# Bright Fluorophores in the Second Near-Infrared Window: HgSe/CdSe Quantum Dots

Ananth Kamath[1], Richard D. Schaller[2,3], and Philippe Guyot-Sionnest[1]*

1: Department of Chemistry and the James Franck Institute, The University of Chicago, 929 East 57th Street, Chicago, Illinois 60637, United States

2: Department of Chemistry, Northwestern University, Evanston, Illinois 60208, United States

3: Center for Nanoscale Materials, Argonne National Laboratory, Lemont, Illinois 60439, United States

**ABSTRACT:** Fluorophores emitting in the NIR-IIb wavelength range (1.5 $\mu$m – 1.7 $\mu$m) show great potential for bioimaging due to their large tissue penetration. However, current fluorophores suffer from poor emission with quantum yields ~2% in aqueous solvents. In this work, we report the synthesis of HgSe/CdSe core/shell quantum dots emitting at 1.7 $\mu$m through the interband transition. Growth of a thick shell led to a drastic increase in the photoluminescence quantum yield, with a value of 55% in nonpolar solvents. The quantum yields of our QDs and other reported QDs are explained well by a model of Forster resonance energy transfer to ligands and solvent molecules. The model predicts a quantum yield >6% when these HgSe/CdSe QDs are solubilized in water. Our work demonstrates the importance of a thick type-I shell to obtain bright emission in the NIR-IIb region.

Fluorophores in the second near-infrared (NIR-II) (1.3 $\mu$m – 2.3 $\mu$m) range have shown potential for non-invasive bioimaging due to reduced autofluorescence, and significantly larger penetration into tissue compared to visible wavelengths.[1–5] The optimal wavelength for in-vivo imaging is the NIR-IIb range (1.5 $\mu$m – 1.7 $\mu$m) which allows deepest penetration as a balance between reduced scattering and absorption by water.[1] In-vivo imaging in the NIR-II region has been demonstrated in imaging mouse brain vasculature,[6] lymph nodes[1] and tumors.[2]

Several fluorophores have been developed for NIR-II bioimaging including organic molecules,[7,8] inorganic quantum dots[1–3] and carbon nanotubes.[6,9] Despite significant progress, organic molecules show very low PLQYs in the NIR, due to their intramolecular vibrational relaxation to molecular vibrations by the energy gap law.[10,11] The brightest NIR fluorophores are based on inorganic quantum dots (QDs) which show QYs up to 43% at 1.5 $\mu$m emission.[12] However, the emission is quenched drastically when the QDs are solubilized in water, with QYs of ~2% at 1.7 $\mu$m. Design of brighter QD fluorophores can lead to deeper penetration for in-vivo imaging, at lower excitation powers.[1]

The brightest reported QDs in the NIR-II range (1.3 $\mu$m – 2.3 $\mu$m) include QD structures based on PbS,[12–16] PbSe,[15] Ag$_2$S,[17] InAs,[2,18] HgTe[19–22] and HgSe.[23,24] Nonradiative relaxation in QDs via surface trapping dominate role in visible-gap QDs, while QDs in the infrared appear to be limited by a near-field Forster energy transfer to ligand molecules on the QD surface.[15,25] This is exacerbated on making the QDs water-soluble, as water is strongly absorbing, and leads to a drastic decrease in the PLQY.[2]

The energy transfer can be suppressed by exchanging the organic ligands by less absorbing molecules, or by growth of a thick type-I shell.[24–26] Solid-state ligand exchange has been demonstrated in Ag$_2$S[17] and PbS[13,14] QD films, which show a PLQY of up to 85% at 1.4 $\mu$m. For in-vivo imaging, the only option is the growth of a thick shell, to suppress energy transfer to the surface ligand and solvent molecules. While thick CdS and CdSe shells have been grown on PbS[12] and PbSe[27] QDs, no substantial improvement in the PLQY is seen. Growth of the thick shell is also accompanied by a redshift of the exciton and lengthening of the PL lifetime,[27] which has been attributed to the quasi-type-II band alignment between the core and the shell.[12,27] Similar observations have been reported for InAs/CdSe QDs, where the PLQY does not improve with shell thickness due to the unfavorable band alignment.[2,28] This motivates the design of a core/shell material system with a type-I alignment, to obtain bright fluorophores in the NIR-II region.

HgSe/CdS QDs show bright mid-infrared emission when n-doping through the intraband transition.[24] Undoped HgSe QDs show emission in the NIR-II range. In contrast to PbS,[12] PbSe[27] and InAs,[2] HgSe has a type-I band alignment with CdS and CdSe,[24,29] which makes it a promising candidate as bright NIR-II fluorophores. CdSe is the ideal shell candidate for obtaining bright emission in HgSe, due to the type-I band alignment and near-zero lattice mismatch. Previous works have reported growth of a thin CdSe shell <~ 1.5nm thickness,[23,29] but growth of a thick spherical shell was not achieved.

In this work, we show the growth of a thick, roughly spherical CdSe shell around HgSe QDs, up to a final QD diameter of 13 nm. These QDs exhibit a PLQY of 55% at 1.7 $\mu$m wavelength, which is ~2-3 times brighter than previous reports. Our calculations show that the QYs are consistent with a Forster energy transfer to ligands and solvent molecules, and demonstrate a strategy to obtain bright NIR-II fluorophores.

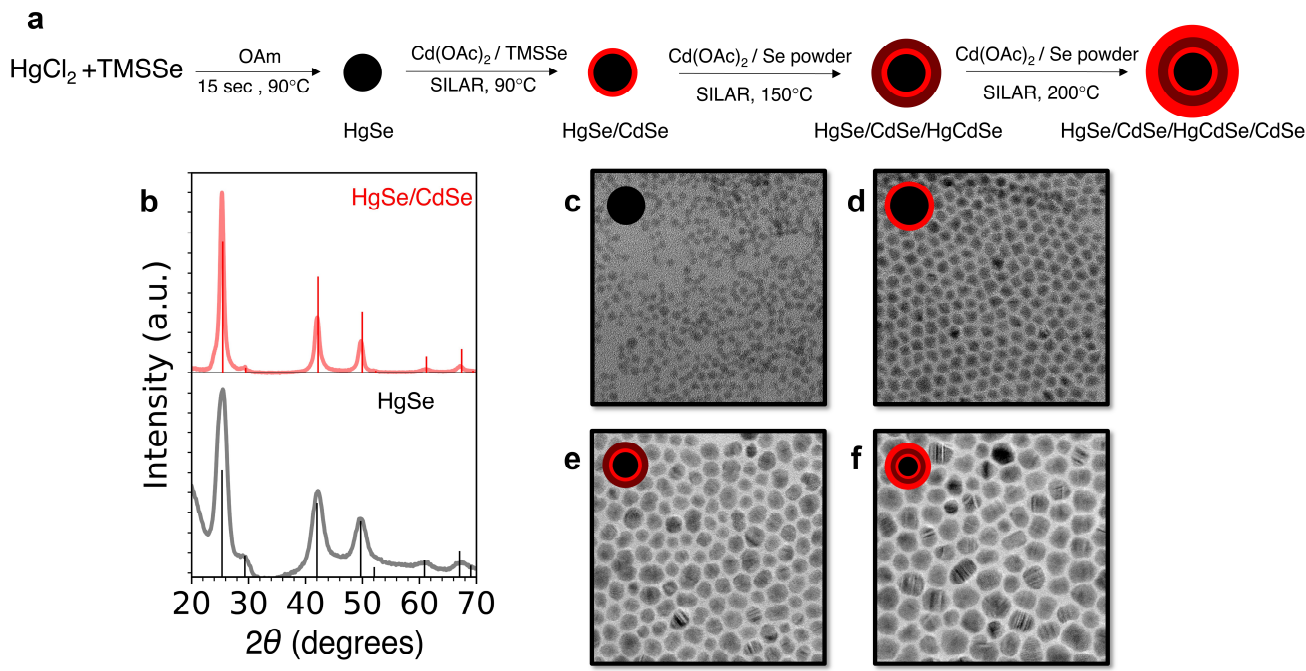

**Figure 1 | Synthesis of thick shell HgSe/CdSe QDs. a**, Schematic of synthesis of HgSe/CdSe quantum dots (QDs). HgSe cores were synthesized by hot injection of $HgCl_2$ and bis(trimethylsilyl)selenide (TMSSe) at 90°C. A thin CdSe shell was then overcoated at 90°C using cadmium acetate and TMSSe, by successive ion layer addition and reaction (SILAR). Subsequent shell growth was performed at 150°C using cadmium acetate and selenium power. At this temperature, around 40-60% of the QDs dissolved and overcoated an alloyed HgCdSe shell around the remaining QDs. The final layers of CdSe were grown at 200°C. **b**, X-Ray diffraction (XRD) spectra of 5.2 nm HgSe (black curve) and 15.1 nm HgSe/CdSe (red curve) QDs, along with calculated bulk XRD spectra for zincblende HgSe (black bars) and CdSe (red bars) with lattice constants 6.08 Å and 6.05 Å respectively. The spectra confirm the growth of CdSe along the zincblende crystal structure like the cores, and the near-zero lattice mismatch between core and shell. **c–f**, Transmission electron microscopy (TEM) images of HgSe and HgSe/CdSe QDs with a diameter of (**c**) 5.2 nm, (**d**) 7.0 nm, (**e**) 10.2 nm, and (**f**) 13.0 nm.

## Synthesis of HgSe/CdSe core/shell QDs

HgSe cores were synthesized by adapting a previous report.[30] The sizes of the QDs were characterized by Transmission electron microscopy (TEM) and Small Angle X-Ray Scattering (SAXS). The cores had a diameter of 4.8nm – 5.2nm for emission at (1.7 µm – 2.0 µm) wavelengths. Most reports of CdSe growth require temperatures in excess of 200°C to obtain large sizes.[27,31] Since HgSe QDs have a poor thermal stability above 100°C, it is necessary to first grow a thin protective CdSe shell at a low temperature before growth of a thick shell.[24] While previous studies report growth of thin shell HgSe/CdSe QDs using c-ALD,[23,29] the procedure is tedious, limited to small reaction scales, and is easily susceptible to independent nucleation.

In order to develop a robust and scalable growth of a thin CdSe shell at a low temperature, we designed a shell growth strategy using cadmium acetate and TMSSe (Fig. 1a). These precursors are highly reactive, and form CdSe nanocrystals at temperatures as low as room temperature. By subsequent addition of Cd- and Se- layers by successive ionic layer adsorption and reaction (SILAR) at 90°C, the shell thickness could be precisely controlled (SI section 1C). The synthesis was robust, and could be scaled up to 80mg of HgSe, with little or no observed independent nucleation up to a thickness of 3 monolayers (SI section 1C).

After growth of 3 monolayers, the HgSe/CdSe QDs were thermally stable up to at least 150°C, while the HgSe cores were not stable above 120°C (SI section 1D).

These thin shell HgSe/CdSe QDs were then used as seeds for growth of a thick CdSe shell. Synthesis of large core/shell nanocrystals requires the use of highly-reactive precursors and a high reaction temperature, in order to maintain a spherical shape and avoid independent nucleation.[24,31–33] We adapted the protocol by X Peng and coworkers, where they demonstrated growth of CdSe using cadmium carboxylates and selenium suspension in oleylamine, which decomposed at a low temperature of 140°C.[34]

A major challenge during growth of the CdSe shell is ripening and dissolution of QDs with a thin CdSe shell, which leads to deposition of an alloyed HgCdSe shell on the remaining QDs.[24] While the optical properties of the surviving HgSe cores are not affected, the alloyed HgCdSe shell absorbs part of the excitation light during PL measurements. To minimize ripening and dissolution of the cores (SI section 1G), we performed the shell growth through a two-step temperature increase. The reaction temperature was set to 150°C for two CdSe monolayers, after which the temperature was increased to 200°C or higher for subsequent shell growth (Fig. 1a). During the heat-up to 150°C, typically ~40-60% of the QDs dissolved and deposited an alloyed shell around the remaining QDs (SI section 3A). The thick shell growth temperature was limited by alloying of the core/shell interface. We observed that cores emitting at 2.0 µm retained stability at 220°C, but cores emitting at 1.7 µm were stable only up to 200°C (SI section 1G). By performing reactions with precisely calculated amounts of cadmium acetate and Se, we were able to grow the QDs up to a final diameter of around 15 nm. The shell growth was spherical till a diameter of 13nm, after which independent nucleation of CdSe was observed at the growth temperature of 200°C (SI section 4A). QDs up to a final size of 16.5 nm could be obtained using a growth temperature of 220°C (SI section 1G).

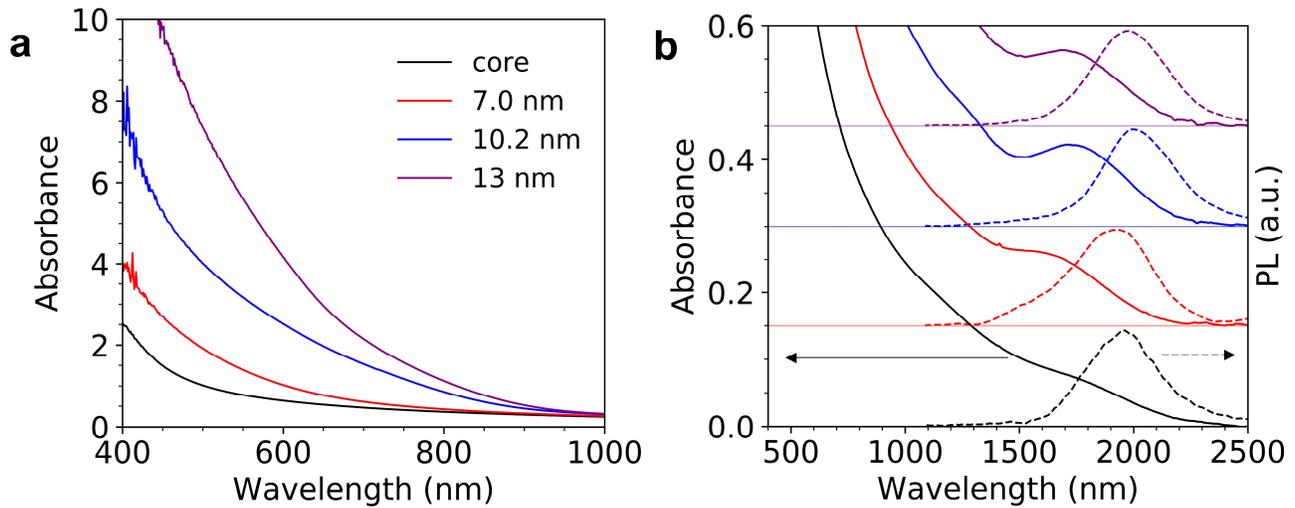

**Figure 2 | Absorption and photoluminescence. a**, Absorption spectra of 5.2 nm HgSe cores and HgSe/CdSe QDs with different sizes. All spectra are normalized at 1400 nm. The increasing band-edge absorption from the CdSe shell is evident from the absorption onset at 900 nm. **b**, Absorption (solid) and photoluminescence (black) (PL) spectra of HgSe and HgSe/CdSe QDs. The absorptions have the same scale as in (**a**), but are vertically shifted for clarity. The absorption and PL spectra show negligible shift on shell growth, confirming the type-I band alignment of HgSe and CdSe.

## Absorption spectra and optical properties of HgSe

An aliquot was taken after synthesis of HgSe cores at 15 seconds. The cores showed both interband absorption at 1.7 µm and an intraband absorption at 4.8 µm, which indicates they are partially n-doped (SI section 4A). By subtracting spectra of the cores at different dopings, we were able to calculate the average doping of the cores to be ~1.05 electrons/QD (SI section 3B).

Aliquots were taken at different stages of the CdSe shell growth. Growth of the thick shell led to a steady increase in the CdSe absorption with a band-edge onset at 900 nm (Fig. 2a). X-Ray Diffraction measurements were performed on films of HgSe and HgSe/CdSe QDs (Fig. 1b). The measured spectra show good agreement with simulated spectra for zincblende HgSe and CdSe. The Scherer sizes calculated for the (220) peak at 42° were 3.7 nm and 9.3 nm respectively (SI section 2E), which show fair agreement with the sizes measured by SAXS.

The HgSe/CdSe QDs were intrinsic and showed a negligible intraband transition, regardless of whether the final layer was Cd- or Se- (SI section 1F). The thickest shell sample was an anomaly, which displayed an intraband absorption with roughly ~0.85 electrons/QD upon ending with a Cd-step (SI section 3B). The interband emission was much weaker after stopping at a Se- step (SI section 1F), so all measurements were performed with Cd- as the final layer.

As seen from the absorption spectrum in Fig. 2b, the absorption features of the HgSe interband transition were maintained on growth of the thick CdSe shell. During growth of the thick shell at 150°C (Fig. 2a, blue), there was a significant decrease in the HgSe absorption intensity (SI section 3A), and a redshift of the absorption peak. This is consistent with dissolution of nearly half of the initial HgSe QDs, and deposition of an alloyed shell around the remaining QDs. The redshift of the absorption suggests that thin shell HgSe/CdSe QDs with a smaller sized core were preferentially dissolved. Subsequent growth of the CdSe shell did not result in a change in the absorption peak wavelength or intensity. The lack of peak shifting confirms the type-I band alignment between HgSe and CdSe. This is in contrast to PbS, PbSe, and InAs QDs, which show a continuous redshift in the exciton peak upon growth of a thick shell due to their quasi type-II band alignment.[2,12,27]

## Photoluminescence and lifetime measurements

Photoluminescence (PL) spectra of the HgSe/CdSe QDs were recorded using an excitation wavelength of 808 nm and an incident power of 15mW, using a step-scan FTIR. All spectra were recorded in solution in tetrachloroethylene (TCE). The HgSe/CdSe QDs showed bright interband emission with no shift in the spectra on shell growth. There was a ~250 nm Stokes shift in the PL (~90 meV), which likely arises from the size distribution of the cores.

Absolute PL quantum efficiency measurements were performed using a Spectralon integrating sphere, with an 808 nm excitation and a PbSe detector. We want to distinguish the PL quantum efficiency (PLQE) from quantum yield (PLQY). When the sample has a partial n-doping, there is an ensemble of QDs with 0, 1 and 2 electrons in the $1S_e$ state (denoted $1S_e(0)$, $1S_e(1)$, and $1S_e(2)$ respectively), each of which would have a specific PLQY. The absolute PLQE can be measured, but the PLQY would have to be calculated from the distribution of QDs with different $1S_e$ occupancies.[24] Only $1S_e(0)$ QDs would be expected to show interband emission due to Kasha's rule. To calculate the PLQY, we estimate the fraction of $1S_e(0)$ QDs in the ensemble by assuming a binomial distribution, and divide the PLQE by this fraction (SI section 3C).[24] Since only the cores and 15.1 nm HgSe/CdSe QDs were partially n-doped, the PLQE and PLQY were the same for QDs of the remaining sizes (SI section 4A).

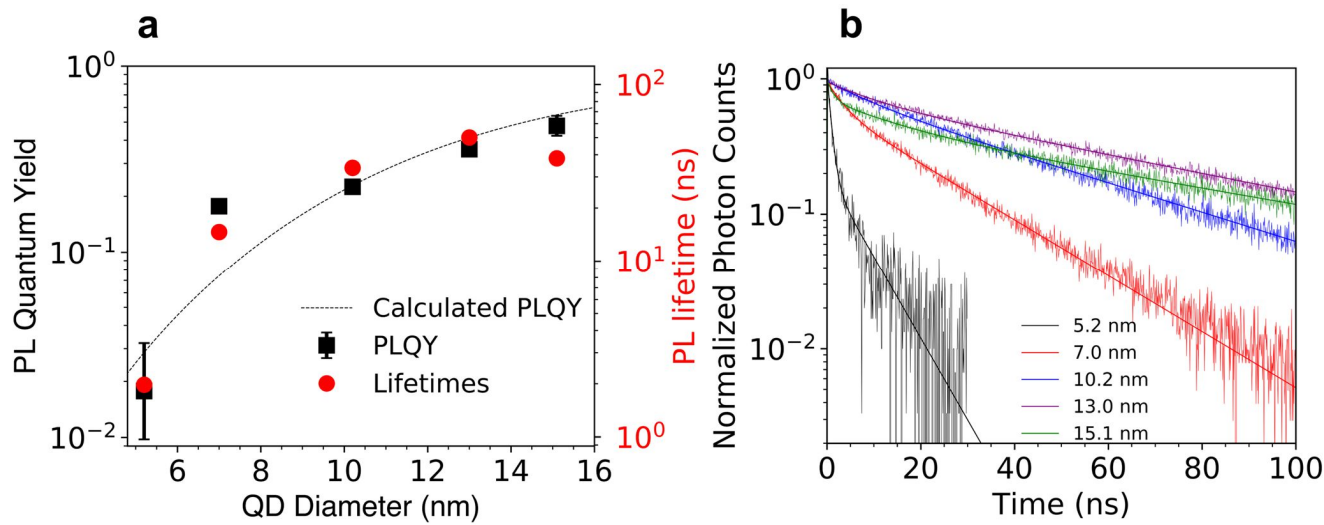

**Figure 3 | Quantum yield and lifetime measurements. a**, Measured photoluminescence quantum yield (black squares) of HgSe/CdSe QDs with different shell thicknesses, calculated quantum yield (dashed curve) modelled by Forster-energy transfer to oleylamine ligands on the QD surface, and measured average PL lifetimes (red circles). The increase in average lifetime with QD size shows a quantitative agreement with the PLQY measurements. Error bars in PLQY correspond to lower and upper limits from the doping estimation. **b**, PL lifetime traces (line scatter) of the same HgSe and HgSe/CdSe QDs in (**a**). The 15.1 nm sample was fit to a triexponential, while the remaining samples were fit to biexponentials (fits are overlayed).

The HgSe cores showed a PLQY of 1.5% at 2.0 μm emission. Growth of the CdSe shell led to a monotonous increase in the PLQY, with a value of 48% for the thickest shell with diameter of 15.1 nm (Fig. 3a). A similar trend was observed for QDs with smaller cores emitting at 1.7 μm, which showed a PLQY of 55% for HgSe/CdSe with a diameter of 12.5 nm (SI section 4B). These QDs are around 2 – 3 times brighter than the brightest fluorophores with PLQY ~20% at 1.7 μm emission.[22,23]

To confirm that the increase in the PLQY is indeed due to a slowing of the nonradiative rate, we measured the PL lifetimes of HgSe and HgSe/CdSe QDs in solution (Fig. 3b). With the exception of the 15.1 nm diameter sample, the lifetime traces of all samples fit well to a biexponential function. The fast component comprised of ~25% of the intensity for the HgSe/CdSe QDs, while it was ~80% for the cores, which is likely due to a larger fraction of trap states in the cores. The 15.1 nm diameter sample was fit to a triexponential with ~30% intensity exhibiting a very fast ~1.4 ns decay. We attribute this to defects arising from the independent nucleation of CdSe on the surface of the QDs (SI section 4A).

We calculated the average PL lifetime to be a harmonic mean of the decay components (SI Section 4A, 4B). The trend of the effective PL lifetime showed quantitative agreement with the PLQY measurements (Fig. 3a). From the measured lifetimes and PLQYs, we calculated the radiative lifetime for each QD sample. The radiative lifetime was similar for all QD samples, giving a value of 127 ns ± 30 ns. The constant radiative lifetime further confirms the type-I alignment of HgSe with CdSe.

## Nonradiative mechanism

Quantum dots in the infrared show PLQYs that are lower than their visible counterparts. To determine the origin of the poor emission, we tabulated the PLQYs of the brightest reported QDs at different wavelengths (Fig. 4b, Data in Supplementary Files). We have plotted the PLQY of our HgSe/CdSe QDs emitting at different frequencies (Fig 4b, green stars). The QDs are 2 – 3 times brighter than previous QDs in the 1.7 μm – 2.0 μm range.

Despite the vast differences in the core and shell materials, all QDs in organic solvents (black and grey points) show a similar decreasing trend in the PLQY as the frequency is reduced. A similar trend is observed for QDs solubilized in water (Fig. 4b, red squares), which have ~5 times lower PLQYs than their counterparts in organic solvents. This general trend suggests that the PLQY is limited by a mechanism that depends on the emission frequency, but is relatively insensitive to the QD material.

Guyot-Sionnest and coworkers[25] proposed a model for nonradiative relaxation involving Forster resonance energy transfer (FRET) of the QD emission to absorption of ligands on the QD surface. This model has been used to explain the lifetime and PLQY of QDs in the NIR and mid-IR regions.[15,19,24–26,36] In this model, the PLQY of QDs in nonpolar solvents is calculated as follows[36] (See SI section 3D):

$$\frac{\gamma_{NR}}{\gamma_R} = \frac{1}{32\,\pi^3 n} \left(\frac{1}{R^3} - \frac{1}{(R+\Delta R)^3}\right) \int \frac{\epsilon''(\bar{v}) f_D(\bar{v})}{\bar{v}^3} d\bar{v}$$

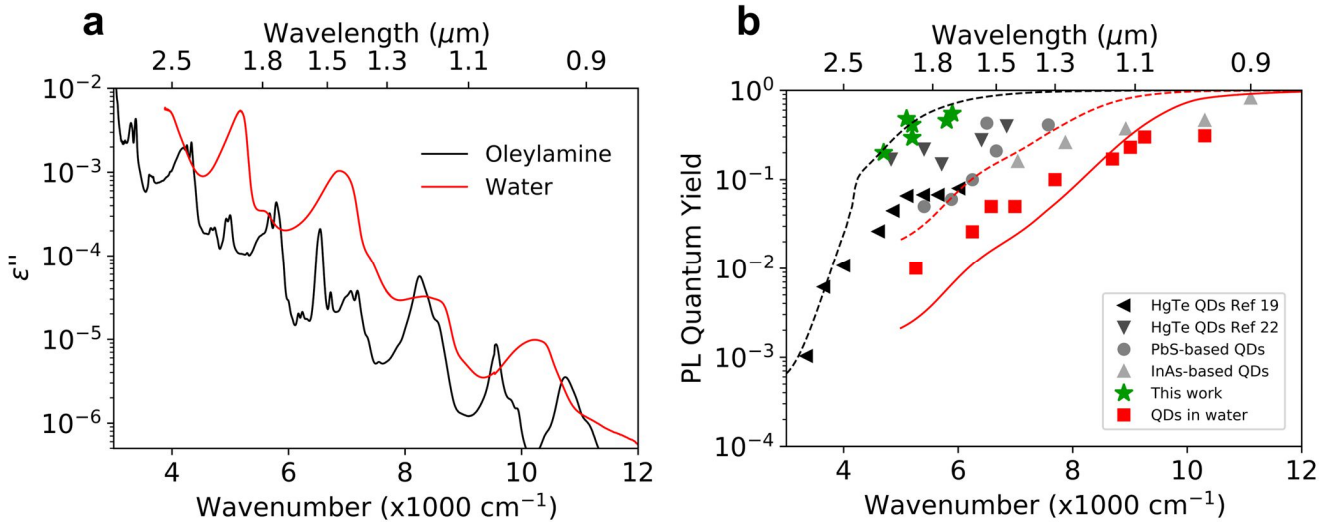

**Figure 4 | Predictions of FRET model, and comparison of PLQYs to other reports. a**, Measured imaginary refractive index of oleylamine and water at a function of frequency. Absorption by water is nearly 5 times stronger than oleylamine. **b**, Compilation of PLQYs of the brightest reported QDs at different emission wavelengths. Black and grey points are QDs in organic solvents.[2,12,15,19,22] Red points are QDs solubilized in water for in-vivo imaging.[1–3,18,35] Green stars are our HgSe/CdSe QDs in nonpolar solvent. The curves are calculated PLQY by FRET to oleylamine ligands (black), and water (red). Solid curve is for 6 nm diameter QDs, and dashed curve is for 13 nm diameter QDs. Ligand absorption and PLQY data are available in supplementary files.

Where $\gamma_{NR}$ is the nonradiative rate, $\gamma_R$ is the radiative rate, $n$ is the index of the solvent, $R$ is the QD radius, $\Delta R$ is the ligand length, $\epsilon''$ is the imaginary dielectric function of the ligand absorption, $f_D$ is the normalized QD emission spectrum and $\bar{v}$ is the frequency in wavenumbers.

The PLQY can be calculated as:

$$\text{PLQY} = \frac{\gamma_R}{\gamma_R + \gamma_{NR}} = \frac{1}{1 + \frac{\gamma_{NR}}{\gamma_R}}$$

Since the nonradiative rate is proportional to the radiative rate, the PLQY is independent of the radiative rate, and hence insensitive to the QD material. This explains the relative insensitivity of the PLQY to the QD material. A notable exception is the series of HgTe QDs reported by Rogach and coworkers,[22] where the PLQY is higher and shows a weaker frequency dependence than other reports.

To calculate the FRET rate for QDs in nonpolar solvents and in water, we measured the absorption spectrum of oleylamine and water and calculated the imaginary index (Fig. 4a). The absorption by water is nearly 5 times stronger than oleylamine. Though the QDs reports in Fig. 4b have different nonpolar ligands, the FRET rate should be similar as the absorption in the NIR-II is dominated by C-H overtones (SI Section 3C). We used the reported surface coverage of oleylamine on the QD surface as 1.8 nm⁻² (see SI Section 3D)[37], and set the emission spectra as Gaussians with standard deviation 0.1 times the emission frequency. We calculated the FRET-limited PLQY as a function of QD diameter (Fig. 3a), and as a function of frequency by setting the QD diameter at 13 nm, which is the average size of the HgSe/CdSe QDs (Fig 4b dashed black curve). The calculations show a good agreement with our measured PLQYs. The trend of calculated PLQY with frequency shows a reasonable agreement with the experimental data, with the scatter likely due to a variation in QD diameters, emission widths, different surface ligands, and contributions from other nonradiative processes.

FRET to water is expected to play a major role for QDs optimized for in-vivo imaging. Water exhibits an absorption in the NIR-II that is around 5 times stronger than oleylamine, and leads to much poorer PLQYs. When we performed the FRET calculation for 6 nm diameter QDs, which is a typical size for water-solubilized QDs (Fig. 4b, solid red curve), we find a fair agreement with the reported PLQYs. Our calculation for 13 nm QDs in water predict that the PLQY should be ~6% at 1.7 µm emission (Fig. 4b, dashed red curve), which is ~3 times brighter than the brightest fluorophores reported.[1] Growth of 20 nm QDs should lead to a PLQY of ~20% in water. While this is not currently possible with our synthetic protocol, further shell growth can be achieved by using a lower-temperature shell material like an HgCdSe alloy.

## Conclusion

Quantum dots (QDs) emitting in the NIR-IIb range show great potential for in-vivo imaging, but the PLQY of the fluorophores are quite low (~2%). To determine the origin of the fast nonradiative decay, we use a model of Forster resonance energy transfer (FRET) to ligands and solvent molecules. Our calculation accounts for the ubiquitous decrease in the PLQY of QDs as the emission frequency is decreased, and also explains the lower PLQY of QDs solubilized in water. Previous attempts to improve the PLQY have been unsuccessful due to the lack of a type-I core/shell system. In order slow the nonradiative decay, we developed the synthesis of a thick CdSe shell on HgSe QDs. The type-I band alignment leads to a negligible spectral shift with increasing shell thickness. The thick shell HgSe/CdSe QDs show a PLQY of ~50% in the 1.7 $\mu$m – 2.0 $\mu$m range, which is ~2-3 times brighter than the previous reported fluorophores. Our work demonstrates the importance of using thick shell QDs with a type-I band alignment, and motivates the use of these QDs for in-vivo imaging.

## Online content

Methods. Supplementary information (SI) contains detailed synthetic protocols, rationale for synthetic design, characterizations, spectroscopy measurement details, calculations, and raw data for QDs emitting at 1.7 $\mu$m and 2.0 $\mu$m.

## Methods

### Thin shell HgSe/CdSe QD synthesis

The following protocol yielded ~20 mg of HgSe cores. Briefly, $HgCl_2$ (0.1 mmol, 27 mg) was added to a 3-neck flask with 5mL oleylamine. The flask was equipped with a stir bar, rubber sleeve stoppers with a thermocouple attached, and connected to an Argon Schlenk line manifold. The flask was heated at 100°C for ~30-60 minutes. The temperature was then set to 90°C. 0.5mL of 0.2M TMSSe solution was injected swiftly into the flask. At 15 seconds, a calculated volume of 0.2M $Cd(OAc)_2$ solution ("0Cd") was injected over a period of ~15 seconds. After 5 minutes, a calculated volume of 0.2M TMSSe solution ("1Se") was injected over a period of ~15 seconds. Subsequent cycles were performed as necessary, with a 2-minute reaction time for Se- cycles and 5-minute reaction time for Cd- cycles. Aliquots were taken if necessary, using a glass syringe with a metal cannula. The reaction was typically stopped after 3 CdSe monolayers, leading to QDs with a diameter ~ 7 nm (denoted HgSe/3CdSe).

The stock solution was transferred into a glass vial and stored in a freezer. The stock solution was directly used for synthesis of a thick CdSe shell.

### Thick shell HgSe/CdSe QD synthesis

The following protocol started with HgSe/3CdSe QDs containing 8 mg of HgSe. The fraction of QDs dissolved during the initial heat-up step was sensitive to the reaction scale.

HgSe/3CdSe stock solution from the freezer was fully thawed. A measured volume of the solution (containing 8 mg of HgSe) was added to a 3-neck flask, equipped with a stir bar, rubber sleeves, a thermocouple, and connected to a Schlenk line. The solution was heated to 150°C. A certain fraction (typically ~60%) of the QDs dissolved during the heat-up stage, and deposited as an alloyed HgCdSe on the remaining QDs.

The Cd- and Se- precursor solutions were kept ready with 1 mL syringes. The Se- suspension was kept on the sonicator, and mixed vigorously before adding the desired volume of precursor. On reaching the set temperature, the Se precursor was added, and left to react for 5 minutes. The Cd precursor was then added, and left to react for 2 minutes. At this stage, the QD diameter was typically ~10.5 nm. One more Se- and Cd- cycle was performed. After the Cd addition, the temperature was set to 200°C. Further Se- and Cd- cycles were performed, with a reaction time of 2 minutes for each cycle. Aliquots were taken if necessary, using a glass syringe with a metal cannula. The QDs diameters increased by roughly 0.7 nm per SeCd cycle. After ~13 nm, the QDs start to develop a tetrahedral-like morphology with a possibility of independent nucleation unless the reaction temperature is increased.

All samples were purified by two cycles of precipitation / dissolution using IPA and TCE.

### Particle size characterizations

Transmission Electron Microscopy (TEM) images were recorded using an FEI Spirit 120kV electron microscope and an FEI Tecnai F30 300kV microscope.

Small Angle X-Ray Scattering (SAXS) measurements were performed using a SAXSLAB GANESHA instrument. The sample was prepared in a Kapton capillary tube and sealed.

X-Ray Diffraction (XRD) measurements were performed using a Rigaku Miniflex Benchtop spectrometer. Samples were dropcasted on a silicon holder.

### Absorption and PL measurements

UV-NIR absorption measurements in the 300 nm – 2500 nm range were performed using an Agilent Cary 5000 UMA Spectrophotometer. Samples were prepared in a glass cuvette with 1 cm path length, using TCE as the solvent.

FTIR absorption measurements in the 1600 nm – 10,000 nm range were performed using a ThermoNicolet iS50 spectrometer. Samples were prepared in a cell with $CaF_2$ windows with 0.5 mm path length, using TCE as the solvent. FTIR measurements allowed a quantitative determination of the amount of HgSe cores in aliquots.

Photoluminescence (PL) spectra were recorded using a step-scan FTIR spectrometer with an MCT detector and a gated integrator. The samples were excited with a 15mW 808nm diode laser, modulated at 90 kHz. A Si wafer was placed in front of the detector to block the excitation light. The transmittance of the solution at 808nm was measured using a Si diode detector behind the sample cell.

Absolute PLQY measurements were performed on QD solutions in TCE in a $CaF_2$ cuvette. The concentration of the solution was adjusted to keep the absorption of the 808 nm light between 20% - 80%, to provide adequate signal, while avoiding reabsorption of the PL (see Fig. S2C-2).

The sample was placed in a Teflon integrating sphere (Thorlabs IS200-4), with the 808 nm laser excitation at one port, with a PbSe detector and Si diode detector at perpendicular ports (see Fig. S2C-1). The 808 nm laser was modulated as a square wave at 1kHz, with average power of 15mW. See SI section 2C for details.

Photoluminescence lifetime measurements were recorded using time-correlated single photon counting for samples in a 1 mm cuvette dispersed in TCE. Samples were excited with a Picoquant 50 ps pulsewidth laser diode operating at 976 nm and 1 MHz repetition rate. PL was collected with a lens, directed thru a silicon longpass filter, dispersed in a 0.3 m spectrograph set to pass the PL emission maximum, and detected with a Quantum Opus superconducting nanowire single photon detector. Single photon arrival times were collected as a histogram for 300 s with a timing bin resolution of 200 ps.

## Data Availability

All supporting data is available in the Supplementary Information. The PLQY data is uploaded as a Supplementary File. Raw data from other measurements is available from the corresponding author upon request.

## Acknowledgements

A.K. is grateful to Aritrajit Gupta for performing SAXS measurements, and to Jennifer Hollingsworth for helpful discussions. A.K. thanks Dmitri Talapin and Bozhi Tian for helpful inputs.

This work is supported by the Department of Energy under award number DE-SC0023210. AK is supported by the Edith Barnard Memorial Fund from the Chemistry Department at the University of Chicago, and by DOE DE-SC0023210. Work performed at the Center for Nanoscale Materials, a U.S. Department of Energy Office of Science User Facility, was supported by the U.S. DOE, Office of Basic Energy Sciences, under Contract No. DE-AC02-06CH11357. This work made use of the shared facilities at the University of Chicago Materials Research Science and Engineering Center, supported by National Science Foundation under award number DMR-2011854; the shared facilities at the University of Chicago Mass Spectrometry Facility, supported by National Science Foundation under award number CHE-1048528; and the University of Chicago electron microscopy facility (RRID:SCR_019198).

## Author Contributions

A.K. and P.G.S conceived the idea. A.K. designed and performed all experiments and analyses. R.D.S. performed the PL lifetime measurements. A.K. wrote the manuscript, with inputs from P.G.S and R.D.S.

## Competing Interest

The authors declare no competing financial interests.

## Additional Information

Supplementary Information is available online.
9

# Supplementary Information: Bright Fluorophores in the Second Near-Infrared Window: HgSe/CdSe Quantum Dots

Ananth Kamath[1], Richard D. Schaller[2,3], and Philippe Guyot-Sionnest[1]*

1: Department of Chemistry and the James Franck Institute, The University of Chicago, 929 East 57th Street, Chicago, Illinois 60637, United States

2: Department of Chemistry, Northwestern University, Evanston, Illinois 60208, United States

3: Center for Nanoscale Materials, Argonne National Laboratory, Lemont, Illinois 60439, United States

# Table of contents

# 1. Syntheses

## 1A: Chemicals

Mercury chloride ($HgCl_2$) was purchased from Alfa Aesar (99.999%, #10808) and Sigma-Aldrich (99.5%, #215465). Mercury bromide ($HgBr_2$, 99%, #83353), cadmium acetate dihydrate ($Cd(OAc)_2$, 98%, #20901), selenium (Se, 99.99%, #229865), oleylamine (OAm, 70%, #909831) dodecanethiol (DDT, 98%, #471364), tetrachloroethylene (TCE, 99%, #443786), ethanedithiol (EdT, 98%, #02390), and octane (99%, #806910) were purchased from Sigma-Aldrich. Bis(trimethylsilyl)selenide (TMSSe, 97%, #SIB1871.0) was purchased from Gelest. Isopropanol (IPA, 99.9%, #A451-4) was purchased from Fisher. (Di-n-dodecyl)dimethylammonium bromide (DDAB, 98%, #B22839) was purchased from Alfa Aesar. TMSSe was stored in a $N_2$ glovebox, while the rest were stored in ambient conditions. All syntheses were performed using the same bottle of oleylamine, as we observed variability in HgSe size and colloidal stability between different batches of the chemical.

## 1B: Precursors and stock solutions

**0.2M TMSSe solution:** The solution was prepared by adding 100 µL TMSSe (0.4 mmol) to 1.9mL oleylamine in a $N_2$ glovebox.

**0.2M $Cd(OAc)_2$ solution:** 78 mg of $Cd(OAc)_2 \cdot 2H_2O$ (0.3 mmol) was added to 1.5 mL oleylamine in a test tube, and heated till the solids dissolved. The solution remained clear as it cooled to room temperature.

**0.1M $Cd(OAc)_2$ solution:** 78 mg of $Cd(OAc)_2 \cdot 2H_2O$ was added to 3 mL oleylamine and prepared as above.

**0.1M Se suspension**: 24 mg selenium powder was added to a vial, and 3 mL oleylamine was added. The suspension was kept on the sonicator during the thick CdSe shell synthesis, and mixed vigorously before use.

**0.1M $HgBr_2$ solution**: 36 mg of $HgBr_2$ (0.1 mmol) was added to 1 mL oleylamine in a test tube, and heated till the solids dissolved. The solution remained clear as it cooled to room temperature.

**0.1M DDAB solution**: 231 mg of DDAB (1 mmol) was added to 5 mL TCE in a vial, and sonicated till the solids dissolved.

**2% EdT/IPA**: 0.1 mL of EdT was added to 5 mL IPA in a vial.

## 1C: Thin shell HgSe/CdSe QD synthesis

**Rationale for synthetic protocol**

HgSe cores were synthesized following the work of Melnychuk et. al.[1] using $HgCl_2$ and bis(trimethylsilyl)selenide (TMSSe) at 90°C – 100°C by hot-injection. A thin CdSe shell had to be grown at a low temperature, in order to impart thermal stability to the QDs before growth of a thicker shell. We did not observe robust and reproducible growth while attempting previous protocols for CdSe shell growth using cALD at room temperature.[2,3] This motivated us to develop a protocol for growth of CdSe around 80°C – 100°C using successive ionic layer adsorption and reaction (SILAR). We chose cadmium acetate ($Cd(OAc)_2$) and bis(trimethylsilyl)selenide (TMSSe) as highly reactive precursors, which formed CdSe at temperatures as low as room temperature.

Due to the labile nature of the HgSe QD surface and the tendency to aggregate after precipitations, we designed the growth of the thin CdSe shell in the HgSe reaction mixture without purification. Unlike typical mercury chalcogenide syntheses which use a 2:1 excess of Hg to chalcogen,[1,4] we used a 1:1 ratio to avoid incorporation of Hg during the CdSe growth. During our trials, we observed that growth of HgSe QDs was complete in less than 15 seconds (Fig. S1C-1), while the n-doping of the QDs increased after further reaction. This suggests that TMSSe completely reacts within 15 seconds, and the HgCl$_2$ in solution continues to react and n-dope the QDs through a surface dipole-dependent Fermi level.[5] This motivated us to start the CdSe growth at 15 seconds after HgSe nucleation.

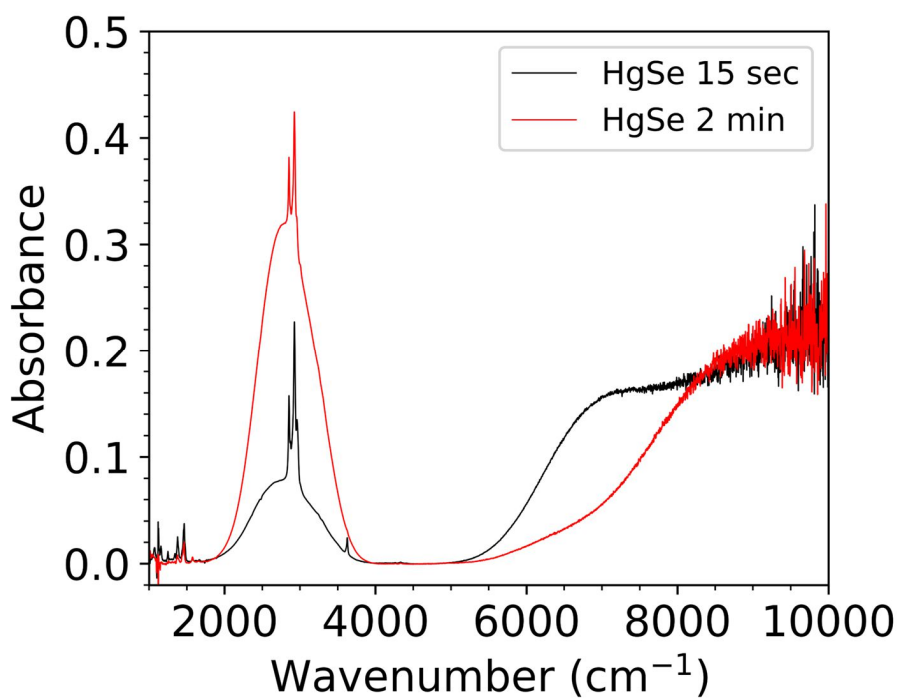

**Fig. S1C-1**: Testing kinetics of HgSe QD synthesis. FTIR absorption spectra of equal volumes of cleaned HgSe aliquots at 15 seconds (black) and 2 minutes (red) after TMSSe injection. The equal absorption at 9000 cm$^{-1}$ and identical intraband peak positions suggests complete reactivity within 15 seconds of TMSSe injection. The n-doping increases through 2 minutes, suggesting the HgCl$_2$ continues to react.

### Calculation of Cd- and Se- precursor volumes

The Cd precursor was 0.2M Cd(OAc)$_2$ in oleylamine, and the Se precursor was 0.2M TMSSe in oleylamine. It was critical to add equal molar amounts of Cd- and Se- precursors to prevent independent nucleation. During synthesis of HgSe,[1] we observed the formation of ~20 mg HgSe (0.07mmol) for addition of 0.1 mmol of TMSSe and 0.2 mmol of HgCl$_2$. This was contrary to the observation of 100% TMSSe reaction at 15 seconds (Fig. S1C-1). We reasoned that the TMSSe stock has a partial reactivity of ~70% due to decomposition over extended storage. The calculated Se- volumes was divided by 0.7 to account for the incomplete reactivity. Precursor volumes for each Cd- and Se- cycle were calculated for saturation of the QDs by one monolayer of the ion. Table S1C-2 shows the calculated volumes for a 20mg scale of HgSe:

| Time (min) | HgSe Mass (mg) | Precursor | Inject vol (mL) | Total vol (mL) | Final Size (nm) |
|---|---|---|---|---|---|
| 0 | | | | 5.5 | 4.8 |
| 0.25 | 20 | 0Cd | 0.17 | 5.67 | 5.15 |
| 5 | 20 | 1Se | 0.27 | 5.94 | 5.5 |
| 7 | 20 | 1Cd | 0.22 | 6.16 | 5.85 |
| 12 | 20 | 2Se | 0.35 | 6.51 | 6.2 |
| 14 | 20 | 2Cd | 0.28 | 6.79 | 6.55 |
| 19 | 20 | 3Se | 0.44 | 7.23 | 6.9 |
| 21 | 20 | 3Cd | 0.34 | 7.57 | 7.25 |

**Table S1C-2**: Calculation of Cd- and Se- precursor volumes during growth of thin shell HgSe/CdSe QDs.

Using this protocol, we were able to observe quantitative growth of the CdSe shell, with roughly ~0.7 nm increase in the total QD diameter as expected. Little or no independent nucleation with observed till 3 monolayers of CdSe. Subsequent layers led to observation of independent nuclei with a smaller final QD size than expected (Fig. S1C-3).

The volumes of injected Cd- and Se- were critical to prevent independent nucleation. If the volumes for each cycle were doubled, substantial independent nucleation was observed.

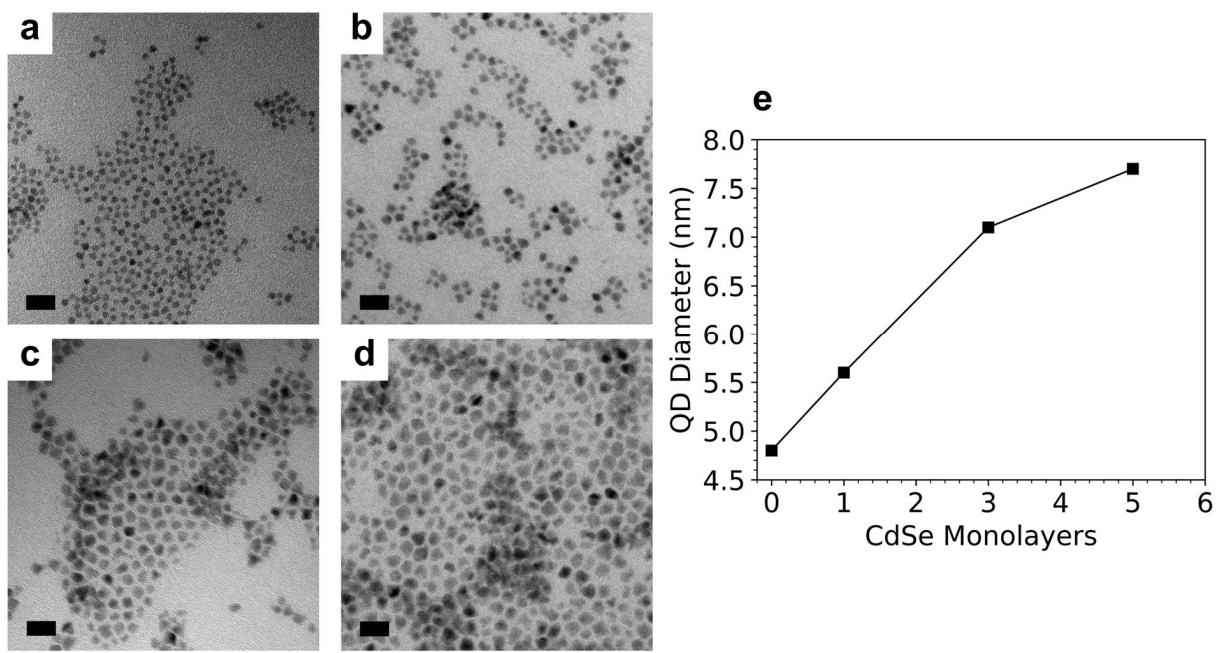

**Fig. S1C-3**: Testing the limit of thin CdSe shell growth. **a-d**, TEM of HgSe QDs (**a**) and HgSe/CdSe QDs with one (**b**), three (**c**) and five (**d**) monolayers of CdSe grown at 90°C. Significant independent nucleation is observed at five monolayers. Scale bars are 20 nm. **e**, Plot of total QD diameter vs CdSe monolayers grown. The size increase is linear till 3 monolayers with roughly 0.7 nm increase per monolayer as expected. The growth tapers at 5 monolayers, likely due to deposition of CdSe precursors on the independent nuclei.

**Synthetic protocol**

The following protocol yielded ~20 mg of HgSe cores. The protocol could be scaled 4x with no noticeable differences. HgSe core size was ~4.8 nm.

$HgCl_2$ (0.1 mmol, 27 mg) was added to a 3-neck flask with 5mL oleylamine. The flask was equipped with a stir bar, rubber sleeve stoppers with a thermocouple attached, and connected to a Schlenk line manifold. Three vacuum (p~1 torr) – Argon flush cycles were performed at room temperature, and the flask was heated at 100°C for ~30-60 minutes. The temperature was then set to 90°C. 0.5mL of 0.2M TMSSe solution was injected swiftly into the flask, which led to the solution instantly turning black, and a stopwatch was started. At 15 seconds, a calculated volume of 0.2M $Cd(OAc)_2$ solution ("0Cd") was injected over a period of ~15 seconds. After 5 minutes, a calculated volume of 0.2M TMSSe solution ("1Se") was injected over a period of ~15 seconds. Subsequent cycles were performed as necessary, with a 2-minute reaction time for Se- cycles and 5 minute reaction time for Cd- cycles. Aliquots were taken if necessary, using a glass syringe with a metal cannula. For HgSe (at 15 secs from start of reaction), the aliquot was added to a test tube containing 0.1M $HgBr_2$ – oleylamine solution (equimolar to the amount of Hg in the aliquot). This was found necessary to maintain colloidal stability during precipitations. Aliquots were rinsed with equal volume of octane to collect residual solution in the syringe. The reaction was typically stopped after 3 CdSe monolayers, leading to QDs with a diameter ~ 7 nm (denoted HgSe/3CdSe).

Concentration of the thin shell HgSe/CdSe stock solution was determined by measuring the absorbance at 700 nm. Absorption at this wavelength should be dominated by the HgSe cores, with negligible shell absorption. The concentration of HgSe was determined using the cross-section of HgSe as $3.5 \times 10^{-18}$ $cm^2$ per Hg atom (calculated from the reported cross-section at 415 nm and the measured HgSe absorption spectrum)[1]. The typical reaction yield was ~20-25 mg of HgSe.

The stock solution was transferred into a glass vial and stored in a freezer. The stock solution was directly used for synthesis of a thick CdSe shell.

Aliquots were purified by two cycles of precipitation – dissolution before characterization. All purifications were performed in air at ambient conditions. The solution was centrifuged to precipitate larger QDs (if any). The solution was then precipitated by addition of IPA (ethanol for HgSe) and centrifugation. The supernatant was discarded, and the precipitate was dispersed in TCE, leading to a concentration of ~10 mg/mL of HgSe. 0.1M DDAB/TCE was added (0.1× volume of TCE), and the solution was precipitated by addition of IPA and centrifugation. The supernatant was discarded, the precipitate was lightly dried by blowing nitrogen, and then dispersed in TCE for characterization.

### 1D: Testing the thermal stability of thin shell HgSe/CdSe QDs

The thin shell HgSe/CdSe QDs were tested for thermal stability at 150°C, which would be the growth temperature of the thicker CdSe shell. Thermal stability was tested through PL measurements as it is more convenient than annealing in solution.

For testing thermal stability of the cores, HgSe/Cd QDs (after the '0Cd' cycle, Table S1C-2) were purified twice, dispersed into TCE and dropcasted into a film on an aluminum substrate. The PL of the film was measured. The sample was annealed at 80°C for 2 minutes, and the PL was recorded again. The annealing was performed again at 120°C for 2 minutes, and the PL was recorded.

For HgSe/CdSe films with different shell thicknesses, the solutions were similarly purified twice, dispersed into TCE and dropcasted on aluminum substrates. The films were treated with a 2% solution of EdT in IPA for removing the long chain ligands. The PL was then recorded. Annealing was sequentially performed at 110°C and 150°C for 2 minutes each, and PL spectra were recorded.

The data is presented in Fig. S1D, where the samples are the same as in Fig. S1C-2. The PL spectra clearly show that the cores are not thermally stable at 120°C, while the thermal stability of HgSe/CdSe QDs progressively increases with number of monolayers. The 3-monolayer sample is thermally stable at 150°C, and these QDs were used as seeds for growth of a thicker CdSe shell.

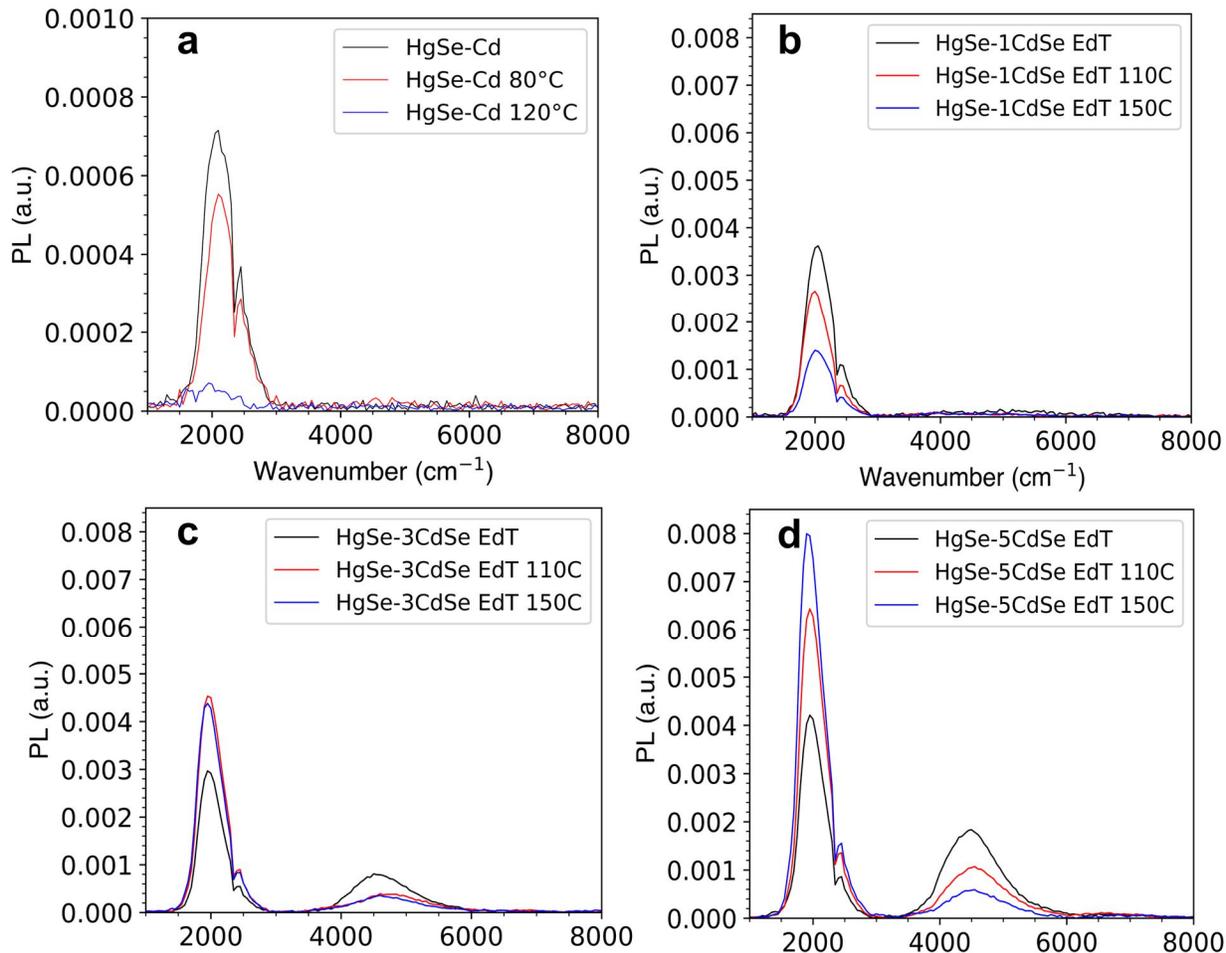

**Fig. S1D**: Testing thermal stability of: **a**, HgSe QDs and **b-d**, HgSe/CdSe QDs with different number of shell monolayers. Samples are the same as in Fig. S1C-2.

### 1E: Thick shell HgSe/CdSe QD synthesis

**Rationale for synthetic protocol**

The thin shell HgSe/CdSe QDs were used as seeds for growth of a thicker shell. We adapted the work of Xiaogang Peng and coworkers, which employed a suspension of selenium in oleylamine, that reacted at temperatures as low as 140°C.[6] We used cadmium acetate as the Cd- reagent, as it is much more reactive than longer carboxylates and enables a lower growth temperature.

The growth temperature was set as 150°C for growth of two monolayers of SeCd. During the heat-up stage, roughly ~60% of the QDs dissolved and deposited as an alloyed HgCdSe shell on the remaining QDs (see section 3A for characterizations). After the first monolayer of SeCd, the QDs were thermally stable, and no further dissolution was observed. After growth of two SeCd layers at 150°C, the temperature was increased to 200°C for further shell growth. The higher growth temperature was necessary to enable growth of a spherical shell. After a final QD size of ~13 nm, the shell started to develop a tetrahedral-like morphology with the onset of independent nucleation (see section 1G). When the growth temperature was 220°C, the QDs remained spherical up to ~16.5 nm, after which independent nucleation was observed. The synthesis was ended with a Cd- cycle, as the PL was much weaker after ending with a Se- cycle (see section 1F).

The thin shell HgSe/CdSe QDs could directly be heated to 200°C/220°C for shell growth, but this led to a larger fraction of dissolved QDs compared to 150°C.

**Calculation of Cd- and Se- precursor volumes**

The Cd precursor was 0.1M Cd(OAc)$_2$ in oleylamine, and the Se precursor was 0.1M Se suspension in oleylamine. Precursor volumes for each Cd- and Se- cycle were calculated for saturation of the QDs by one monolayer of the ion. After growth of the first SeCd monolayer, the calculation was modified to account for the ~60% dissolution of QDs. For example, starting with 7.0 nm diameter HgSe/CdSe QDs with 8 mg of HgSe, the expected QD diameter after one SeCd layer is 7.7 nm, but the measured size is typically 10.5 nm, which corresponds to a 60% dissolution of QDs. The amount of HgSe cores remaining is therefore 4.8 mg. The subsequent volumes of Se- and Cd- additions are calculated to account for this change in size and HgSe amount.

Table S1E is an example of the volumes added for an 8mg scale of HgSe:

| Temperature (°C) | Time (min) | HgSe Mass (mg) | Precursor | Inject vol (mL) | Total vol (mL) | Final Size (nm) |
|---|---|---|---|---|---|---|
| | | | | | 3.10 | 7.00 |
| 150 | 0 | 8 | 4Se | 0.28 | 3.38 | 7.35 |
| 150 | 5 | 8 | 4Cd | 0.31 | 3.69 | 10.45 |
| 150 | 7 | 3.2 | 5Se | 0.25 | 3.93 | 10.80 |
| 150 | 12 | 3.2 | 5Cd | 0.26 | 4.19 | 11.15 |
| 200 | 14 | 3.2 | 6Se | 0.28 | 4.47 | 11.50 |
| 200 | 16 | 3.2 | 6Cd | 0.30 | 4.77 | 11.85 |
| 200 | 18 | 3.2 | 7Se | 0.31 | 5.08 | 12.20 |
| 200 | 20 | 3.2 | 7Cd | 0.33 | 5.41 | 12.55 |
| 200 | 22 | 3.2 | 8Se | 0.35 | 5.77 | 12.90 |
| 200 | 24 | 3.2 | 8Cd | 0.37 | 6.14 | 13.25 |

**Table S1E**: Calculation of Cd- and Se- precursor volumes during growth of thick shell HgSe/CdSe QDs.

**Synthetic protocol**

The following protocol started with HgSe/3CdSe QDs containing 8 mg of HgSe. The fraction of QDs dissolved during the initial heat-up step was sensitive to the reaction scale.

HgSe/3CdSe stock solution from the freezer was fully thawed. A measured volume of the solution (containing 8 mg of HgSe) was added to a 3-neck flask, equipped with a stir bar, rubber sleeves, a thermocouple, and connected to a Schlenk line. The solution was heated to 150°C. A certain fraction (typically ~60%) of the QDs dissolved during the heat-up stage, and deposited as an alloyed HgCdSe on the remaining QDs.

The Cd- and Se- precursor solutions were kept ready with 1 mL syringes. The Se- suspension was kept on the sonicator, and mixed vigorously before adding the desired volume of precursor.

On reaching the set temperature, the Se precursor was added, and left to react for 5 minutes. The Cd precursor was then added, and left to react for 2 minutes. At this stage, the QD diameter was typically ~10.5 nm.

One more Se- and Cd- cycle was performed. After the Cd addition, the temperature was set to 200°C. Further Se- and Cd- cycles were performed, with a reaction time of 2 minutes for each cycle. Aliquots were taken if necessary, using a glass syringe with a metal cannula. Aliquots were rinsed with equal volume of TCE to collect residual solution in the syringe. The QDs diameters increased by roughly 0.7 nm per SeCd cycle. After ~13 nm, the QDs start to develop a tetrahedral-like morphology with a possibility of independent nucleation unless the reaction temperature is increased.

Aliquots were purified by two cycles of precipitation – dissolution before characterization. All purifications were performed in air at ambient conditions. The solution was centrifuged, leading to precipitation of a fraction of the QDs (which increased with the QD size). If QDs remained in the supernatant, 0.1M DDAB/TCE was added (0.1× volume of solution), and the solution was precipitated by addition of IPA and centrifugation. The QDs precipitated in fractions (likely due to broad size distribution), and it was necessary to performed several IPA additions and centrifugations to completely precipitate the QDs. The supernatant was discarded, and the precipitate was dispersed in TCE, leading to a concentration of ~5 mg/mL of HgSe. 0.1M DDAB/TCE was added (0.1× volume of TCE), and the solution was precipitated by addition of IPA and centrifugation. Precipitation of several fractions was observed, and the QDs were fully precipitated. The supernatant was discarded, the precipitate was lightly dried by blowing nitrogen, and then dispersed in TCE for characterization.

### 1F: Influence of Cd- and Se- cycles on doping and PL of HgSe/CdSe QDs

To decide between ending the HgSe/CdSe synthesis with a Cd- or Se- cycle, we tested the effect of the final cycle on the doping and the PL of the QDs. Thick shell HgSe/CdSe QDs with a final diameter of ~10-12 nm were synthesized at 180°C. An aliquot was taking after a Se- cycle. Different amounts of Cd- were then added (corresponding to 1, 2 and 3 surface equivalents of Cd), with aliquots between additions. Partial independent nucleation was observed, but this should not affect the conclusions of these trials.

The QDs were purified twice by precipitation-dissolution, before redispersing in TCE. The final solutions were characterized by FTIR absorption and PL spectroscopy (Fig. S1F). A slight decrease in the n-doping was observed after Cd- treatment, but the PL was much stronger for the Cd- samples compared to the Se- sample. This motivated us to stop the HgSe/CdSe syntheses at the Cd- cycle.

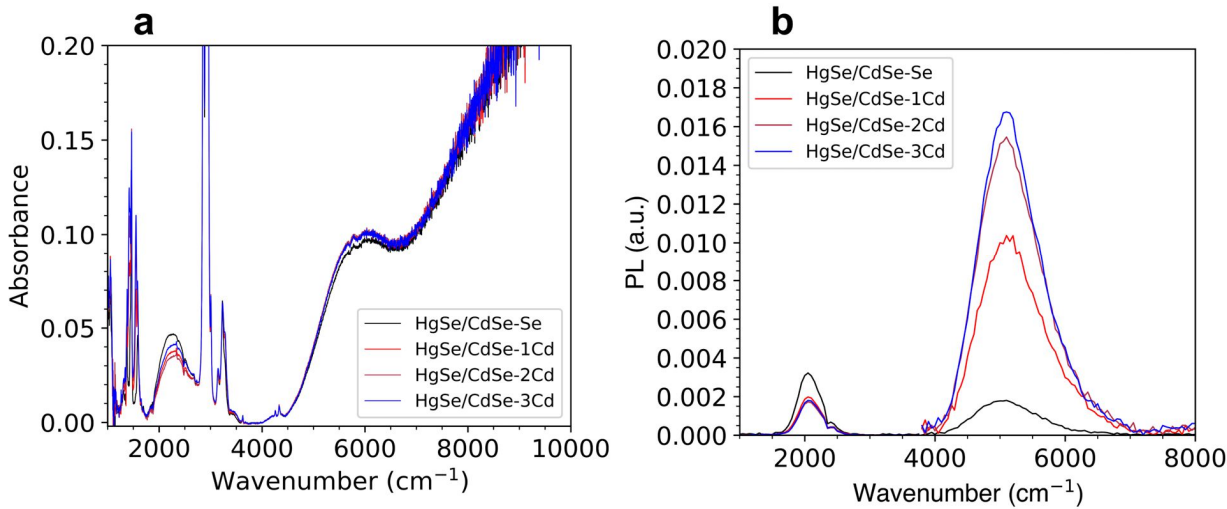

**Fig. S1F**: Effect of Se- and Cd- half cycles on absorption and PL of ~10-12 nm HgSe/CdSe QDs. Testing thermal stability of: **a**, HgSe QDs and **b-d**, HgSe/CdSe QDs with different number of shell monolayers. Samples are the same as in Fig. S1C-2.

## 1G: Effect of synthesis temperature on dissolution, alloying, and morphology of HgSe/CdSe QDs

The synthesis temperature was crucial to the morphology of the HgSe/CdSe QDs. If the temperature was too low, the onset of independent nucleation was observed at smaller QD sizes (Fig. S1G-1). If the temperature was too high, alloying at the core/shell interface was observed, with a blueshift in the absorption and PL (Fig. S1G-2). Smaller sized cores emitting at 5800 cm$^{-1}$ showed core/shell interface alloying at 220°C. We thus set the optimal synthesis temperature as 200°C to achieve large shell thicknesses while avoiding interfacial alloying.

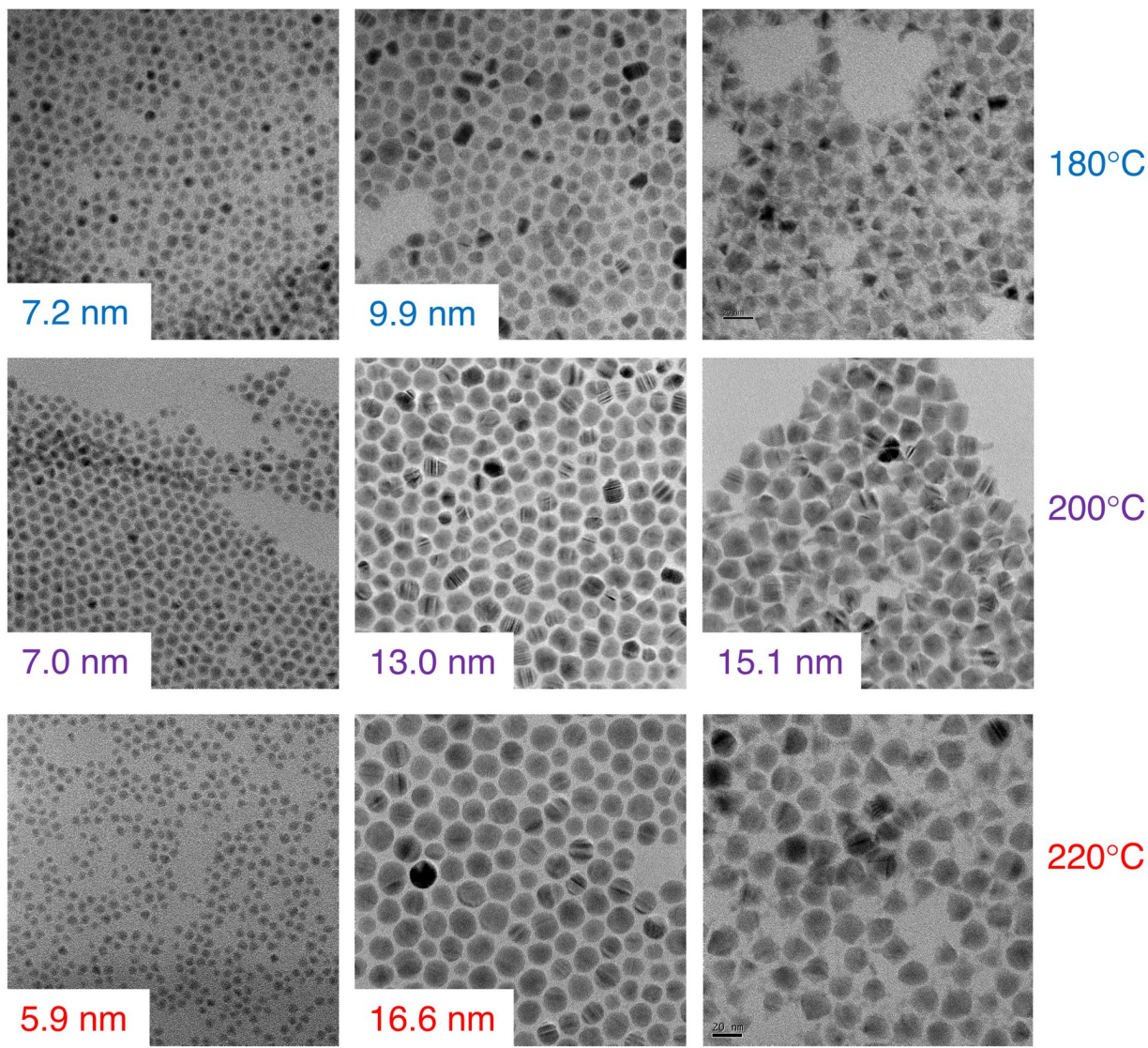

**Fig. S1G-1**: Effect of synthesis temperature on the onset of independent nucleation. The onset of independent nucleation at 180°C, was at a size around 10 nm; at 200°C it was ~15 nm, and at 220°C it was ~17 nm

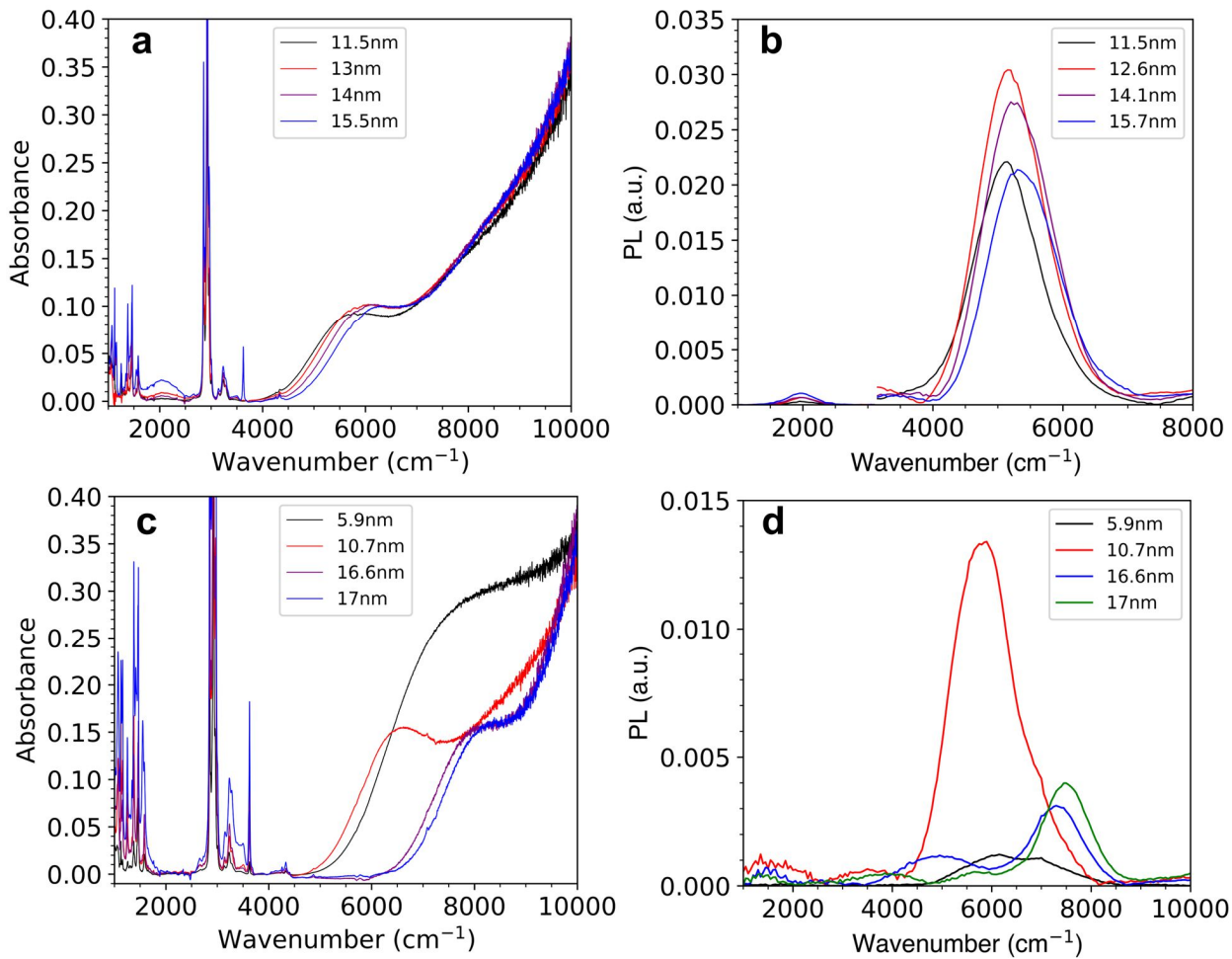

**Fig. S1G-2**: Core/shell interface alloying at 220°C, for HgSe/CdSe QDs emitting at different frequencies. The red curve shows the shell grown at 150°C, and the thicker shell samples are grown at 220°C **a-b**, Absorption and PL of HgSe/CdSe QDs emitting at 5200 cm$^{-1}$. There is a small, but insignificant blueshift in the absorption and PL during subsequent stages of shell growth at 220°C. **c-d**, Absorption and PL of HgSe/CdSe QDs emitting at 5800 cm$^{-1}$. There is a large blueshift in the absorption and PL on heating to 220°C, which indicates alloying at the core/shell interface. This shows that the smaller cores have a lower temperature for alloying at the core/shell interface.

# 2. Characterizations

## 2A: Particle size characterization by TEM and SAXS measurements

Transmission Electron Microscopy (TEM) images were recorded using an FEI Spirit 120kV electron microscope and an FEI Tecnai F30 300kV microscope. Samples were prepared by dropcasting the purified QD solution in TCE on a Formvar/Carbon 200 mesh grid (Ted Pella 01801), and evaporated in a vacuum pump before performing measurements. The QD size was determined using imageJ software, by visually estimating the boundary of ~100 nanoparticles, measured along a single orientation for consistency.

Small Angle X-Ray Scattering (SAXS) measurements were performed using a SAXSLAB GANESHA instrument. The sample was prepared in a Kapton capillary tube and sealed.

## 2B: Absorption measurements

UV-NIR absorption measurements in the 300 nm – 2500 nm range were performed using an Agilent Cary 5000 UMA Spectrophotometer. Samples were prepared in a glass cuvette with 1 cm path length, using TCE as the solvent.

FTIR absorption measurements in the 1600 nm – 10,000 nm range were performed using a ThermoNicolet iS50 spectrometer. Samples were prepared in a cell with $CaF_2$ windows with 0.5 mm path length, using TCE as the solvent. FTIR measurements allowed a quantitative determination of the amount of HgSe cores in aliquots.

## 2C: Photoluminescence and PLQY measurements

**PL spectra**:

Samples were prepared in a cell with CaF2 windows with 0.5 mm path length, using TCE as the solvent. Photoluminescence (PL) spectra were recorded using a step-scan FTIR spectrometer with an MCT detector and a gated integrator. The samples were excited with a 15mW 808nm diode laser, modulated at 90 kHz. A Si wafer was placed in front of the detector to block the excitation light. The transmittance of the solution at 808nm was measured using a Si diode detector behind the sample cell. The spectra were corrected for the combined optics/spectrometer/detector combined response, which was determined by measuring the spectrum of a 1268 Kelvin blackbody at the sample position. The PL spectra were normalized by the fraction of 808 nm light absorbed, and were also corrected for the absorption by TCE. The PLQY could be calculated as the area under the corrected PL spectrum, by comparing to the measured PLQY of a reference sample. Since the spectrometer sensitivity in the 4000 – 8000 $cm^{-1}$ was weak, the PLQYs determined this way had an uncertainty of a factor of ~2.

**Absolute PLQY measurements**:

Setup and sample preparation:

Absolute PLQY measurements were performed on QD solutions in TCE in a $CaF_2$ cuvette. The concentration of the solution was adjusted to keep the absorption of the 808 nm light between 20% - 80%, to provide adequate signal, while avoiding reabsorption of the PL (see Fig. S2C-2). The cuvette area is 0.65 $cm^2$, smaller than the 1.2 $cm^2$ input port in order to minimize indirect re-absorption of incident and emitted light in the integrating sphere.

The sample was placed in a Spectralon integrating sphere (Thorlabs IS200-4), with the 808 nm laser excitation at one port, with a PbSe detector (Thorlabs PDA20H) and Si diode detector at perpendicular ports (see Fig. S2C-1). The 808 nm laser was modulated as a square wave at 1kHz, with average power of 15mW. PLQYs measured at 150mW excitation were roughly ~15% lower than 15mW excitation, likely due to localized heating at the laser spot.

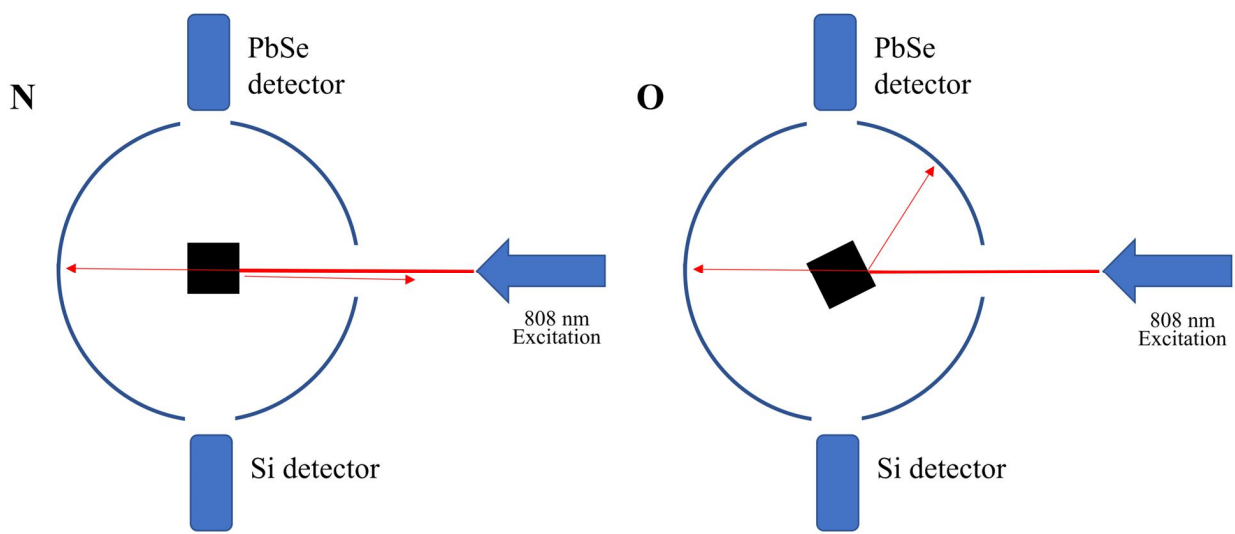

**Fig. S2C-1**: Schematic of PLQY measurements in a Spectralon integrating sphere. The cuvette was oriented in two angles: the 'N' (normal) geometry and the 'O' (oblique) geometry.

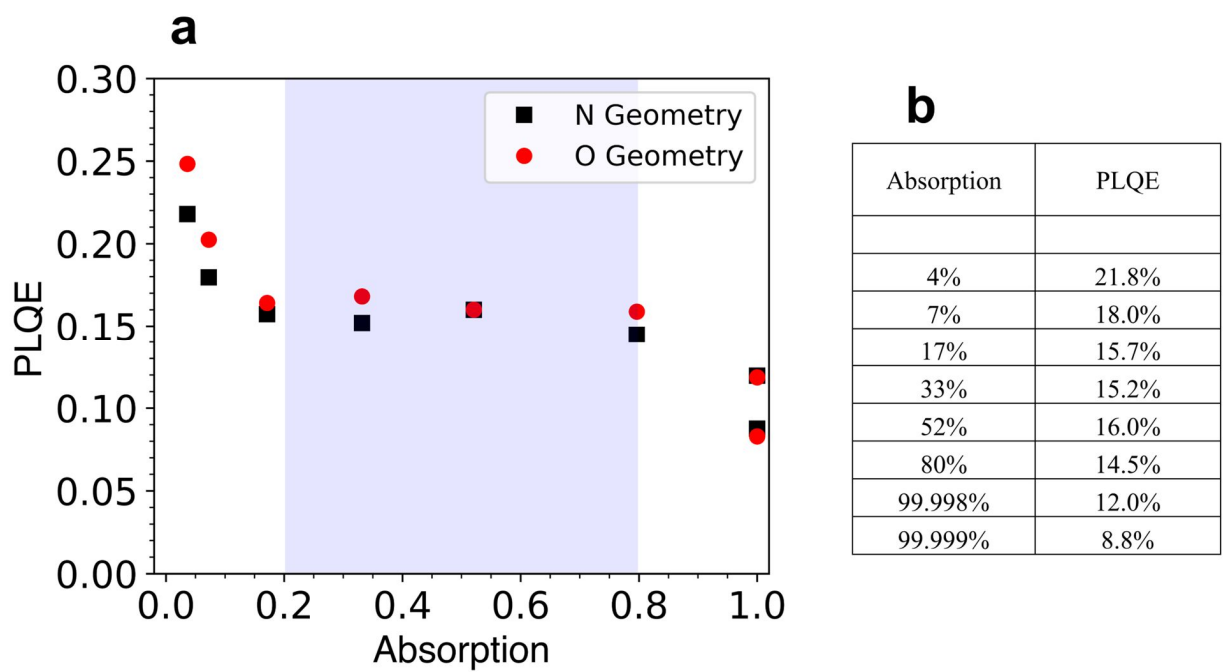

**Fig. S2C-2**: **a**, Effect of the absorption (at 808 nm) on the measured PLQE. At absorption >~ 0.8, the sample reabsorbs the PL light emitted along the excitation direction, and leads to underestimation of the PLQE. At a very high absorption, the measured PLQE saturates to half of the actual value. At absorption <~0.2, the PL signal is low and comparable to the signal from the blank TCE sample, and leads to overestimation of the PLQE. The PLQE measurement is accurate when the absorption is between 0.2 – 0.8, denoted by the shaded region. **b**, Data for PLQE vs absorption for the 'N' geometry.

| Absorption | PLQE |
| --- | --- |
|  |  |
| 4% | 21.8% |
| 7% | 18.0% |
| 17% | 15.7% |
| 33% | 15.2% |
| 52% | 16.0% |
| 80% | 14.5% |
| 99.998% | 12.0% |
| 99.999% | 8.8% |

Direct reflection of the excitation light into the PbSe or Si detectors is avoided as this leads to abnormally high signals. The measurements are performed in two geometries: the 'N' geometry, where the $CaF_2$ cuvette is normal to the excitation laser, and the light reflected off the cuvette is sent out of the integrating sphere. In the 'O' geometry, the cuvette is rotated at an angle, so that the reflected excitation light is incident on a wall of the integrating sphere. To precisely obtain the N geometry, the cuvette is first oriented visually to make the face parallel to the excitation light, and is rotated until the signal on the PbSe detector is at a minimum. To obtain the O geometry, the cuvette is rotated until the signal on the PbSe detector is relatively insensitive to the cuvette orientation. If the cuvette is further rotated, a drastic increase in the PbSe signal is observed, which indicates direct reflection of the excitation light to the detector. By such control of the cuvette orientation, we obtained PLQY measurements with a relative error of only ~3%.

Measurement

A blank TCE solution is placed in the integrating sphere in the N geometry. The signal on the Si detector is $N_{Si}(TCE)$, and the signal on the PbSe detector is $N(TCE)$. A silicon wafer is then placed in front of the PbSe detector to block the 808 nm excitation, giving a signal of $N_{PbSe-B}(TCE)$. Similarly, the signals in the O geometry are $O_{Si}(TCE)$, $O_{PbSe}(TCE)$, and $O_{PbSe-B}(TCE)$ respectively. The signals $N_{PbSe-B}(TCE)$ and $O_{PbSe-B}(TCE)$ should be zero in principle, but a small ~25 mV signal is observed, which sets the lower bound for the PLQY that can be measured with this setup.

The QD solution is then placed in the integrating sphere and measurements are performed in the two geometries. PbSe is a photon detector with a close to linear response (V/W) throughout the 800 nm – 4000 nm range, therefore the signal is taken to be directly proportional to the photon count.

The 'N' and 'O' geometries provided two measurements of the PLQY.

For the N-geometry, the signal from the laser excitation for the clean TCE cell is:

$$Excitation = N_{PbSe}(TCE)$$

This comes from the laser light scattering from the sphere and some of that light getting to the detector.

$N_{PbSe-B}(QD)$ is the signal arising from emission of the sample. Taking into account the transmittance of Si is 56% in the 1.3 μm – 5 μm range, the QD emission signal is:

$$Emission = \frac{N_{PbSe-B}(QD)}{0.56}$$

The QD absorption is determined by subtracting the emission signal from the PbSe signal such that

$$Absorption = 1 - \frac{N_{PbSe}(QD) - \frac{N_{PbSe-B}(QD)}{0.56}}{N_{PbSe}(TCE)}$$

The QD absorption is independently determined from the signal on the Si diode:

$$Absorption = 1 - \frac{N_{Si}(QD)}{N_{Si}(TCE)}$$

The absorptions determined from both measurements were close, with a relative standard deviation of <10%

The PLQY is then determined as:

$$PLQY = \frac{Emission}{Absorption \times Excitation}$$

A similar calculation was done for the 'O' geometry. Measurements in the two independent geometries provided PLQY values that are close, with a relative standard deviation <3%. We used this is as the uncertainty in the absolute PLQY measurement. Additional uncertainty could arise from a nonlinear response curve for the PbSe diode detector.

Raw data is available in supplementary files.

### 2D: Photoluminescence lifetime measurements

Photoluminescence lifetime measurements were recorded using time-correlated single photon counting for samples in a 1 mm cuvette dispersed in TCE. Samples were excited with a Picoquant 50 ps pulsewidth laser diode operating at 976 nm and 1 MHz repetition rate. PL was collected with a lens, directed thru a silicon longpass filter, dispersed in a 0.3 m spectrograph set to pass the PL emission maximum, and detected with a Quantum Opus superconducting nanowire single photon detector. Single photon arrival times were collected as a histogram for 300 s with a timing bin resolution of 200 ps.

### 2E: X-Ray Diffraction Measurements

X-Ray Diffraction (XRD) measurements were performed using a Rigaku Miniflex Benchtop spectrometer. Samples were dropcasted on a silicon holder. Peak fitting was performed using the Rigaku SmartLab Studio II software. The background of the spectrum was calculated as a polynomial, and the peaks were fit to Split pseudo-Voigt functions.

The pXRD spectra for bulk zincblende HgSe and CdSe were simulated using the software *Diamond 3.2*, using lattice constants of 6.08 Å and 6.05 Å respectively.

# 3. Calculations and analyses

## 3A: Dissolution of thin shell HgSe/CdSe QDs and deposition of HgCdSe shell

During the heat-up of thin shell HgSe/CdSe QDs to 150°C, we noticed the measured QD size to be much larger than the calculated size. This was accompanied by a decrease in the HgSe absorption (as determined by quantitative FTIR spectra of aliquots). We attributed this to dissolution of a fraction of the HgSe/CdSe QDs, and deposition on the remaining QDs as a HgCdSe shell. The absorption and PL of the surviving QDs were not affected by the dissolution. Estimation of the dissolution fraction is shown below:

A solution of 7.0 nm thin shell HgSe/CdSe QDs (containing 2mg of HgSe, measured by absorption at 700 nm) was cleaned and the FTIR spectrum was recorded. A reaction mixture of thin shell HgSe/CdSe (containing 8mg of HgSe) was heated to 150°C and one Se-Cd- cycle was performed (see Section 1E). A quarter of the reaction mixture was taken as an aliquot, purified twice, and an FTIR spectrum was recorded.

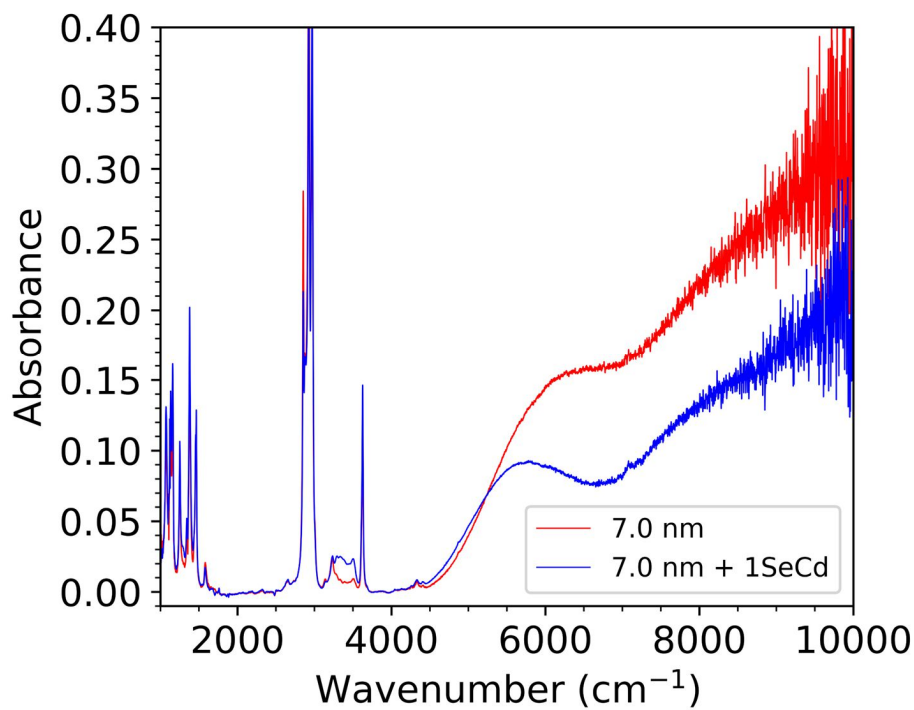

**Fig. S3A**: FTIR absorption spectra of thin shell HgSe/CdSe before (red) and after (blue) performing one Se-Cd cycle. The HgSe absorption decreases by ~41%.

The ratio of absorbance at 8000 cm$^{-1}$ is 0.13 / 0.22 = 0.59, which suggests that 41% of the initial QDs are dissolved.

As an independent measure of the dissolved fraction, we calculated the discrepancy between the measured and calculated TEM size. The expected size after growth of one SeCd cycle is 7.7 nm, while the measured size was 10.2 nm (measured by SAXS). This suggests that the fraction of surviving QDs is $(7.7/10.2)^3 = 0.43$, which suggests that 57% of the initial QDs are dissolved. This estimate qualitatively agrees with the decrease in the intensity of the HgSe absorption.

## 3B: Estimating PLQY of partially doped HgSe and HgSe/CdSe QDs

The HgSe and 15.1 nm HgSe/CdSe QDs were observed to be partially n-doped, as seen from the intraband absorption (Fig. S3B-1) and presence of both interband and intraband emission (Section 4A). It is necessary to determine the doping fraction in order to calculate the interband PLQY. We followed the procedure by Kamath et. al.[7] to measure the doping. The absorption spectrum of the QD sample in TCE was first measured, after which the PL spectrum was recorded. The doping of the sample would typically increase after the PL measurement, likely due to photochemical processes on excitation with the 808 nm laser. The spectra were subtracted to yield the intraband absorption and interband bleach (Fig. S3B-1, blue). This curve was used to extrapolate the HgSe absorption to determine the intraband absorption for a fully n-doped sample.

One way is to add the blue spectrum to the HgSe absorption, till the interband edge appears flat. This gives the lower limit on the intraband absorbance of a fully n-doped sample (Fig. S3B-1, green). Another way is to add the blue spectrum to the HgSe spectrum till the absorbance at the interband peak (6000 cm$^{-1}$) is zero. This gives the upper limit to the absorbance of a fully n-doped sample (Fig. S3B-1, purple). The $1S_e$ occupancy was then calculated to be ratio of the intraband peak absorbance of the HgSe spectrum to the extrapolated spectrum. The lower and upper limit on the $1S_e$ occupancy was determined to be 0.77 and 1.3 electrons/QD respectively. The average doping was thus found to be 1.04 electrons / QD.

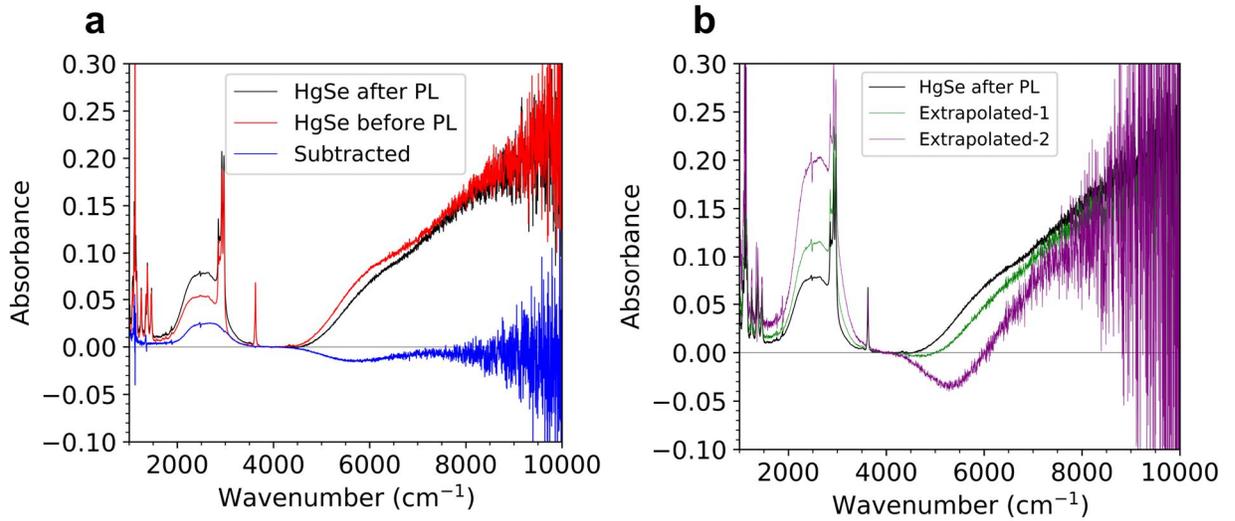

**Fig. S3B-1**: **Estimating the doping of HgSe QDs**. **a**, Absorption spectra of HgSe QDs in TCE before (red) and after (black) PL measurement. The spectra are subtracted (blue) to obtain the intraband absorption and interband bleach. **b**, The blue curve from (**a**) is added to the HgSe spectrum till the interband edge appears to be flat (green), or gives zero absorbance at 6000 cm$^{-1}$ (purple).

Assuming a binomial distribution for QDs with 0, 1 and 2 electrons in the $1S_e$ state (denoted $1S_e(0)$, $1S_e(1)$ and $1S_e(2)$ respectively), the fraction of QDs in $1S_e(0)$ was calculated as:[7]

$$1S_e(0)\ fraction\ =\ \left(1 - \left(\frac{N_e}{2}\right)\right)^2 \qquad (S3B-1)$$

Where $N_e$ is the average number of electrons in $1S_e$. Using $N_e$ = 1.04, we get the average $1S_e(0)$ occupancy is 23%, with lower and upper limits 12% and 38% respectively.

For 15.1 nm HgSe/CdSe, a similar subtraction was performed. Since the doping did not decrease after PL measurement, we used the absorption spectrum of the 13.0 nm sample, which was completely undoped. Since the QD absorption shapes were identical (except for the ligand absorption), we could perform a reliable subtraction using the two different samples (Fig. S3B-2). Similar to the HgSe sample, the blue curve was added to the 15.1 nm HgSe/CdSe QD absorption to determine the lower and upper limits on the absorption of a fully n-doped sample. The lower and upper limit on the $1S_e$ occupancy was determined to be 0.76 and 0.93 electrons/QD respectively. The average doping was thus found to be 0.85 electrons / QD.

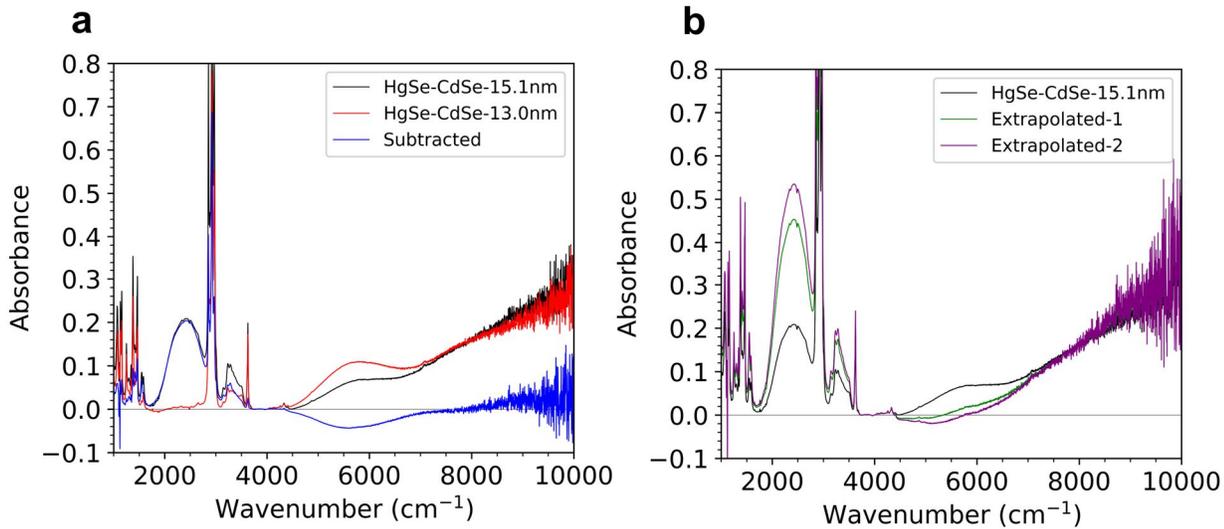

**Fig. S3B-2**: **Estimating the doping of 15.1 nm HgSe/CdSe QDs**. **a**, Absorption spectra of HgSe/CdSe QDs with sizes 15.1 nm (black) and 13.0 nm (red). The spectra are subtracted (blue) to obtain the intraband absorption and interband bleach. **b**, The blue curve from (**a**) is added to the 15.1 nm HgSe/CdSe spectrum till the interband edge appears to be flat (green), or gives zero absorbance at 5800 cm$^{-1}$ (purple).

Using $N_e$ = 0.85 (Eq. S3B-1), we get the average $1S_e(0)$ occupancy is 33%, with lower and upper limits 29% and 38% respectively.

The PLQY was calculated by dividing the measured PLQE (Section 4A, 4B) by 0.23 (for HgSe) and 0.33 (for 15.1 nm HgSe/CdSe). The error bar limits (in Fig. 3a of main text) were set using the lower and upper limits for the $1S_e(0)$ occupancy.

### 3C: Imaginary dielectric function of oleylamine, dodecanethiol and water

The absorption spectra of oleylamine, dodecanethiol and water were measured in the 750 nm – 2500 nm range. The spectra were recorded using TCE as a background.

For oleylamine, spectra were recorded in a 1 cm cuvette. Accurate spectra were obtained at absorbance >~0.01 (due to reflection from index difference of TCE and oleylamine) and <2 (due to spectrometer sensitivity). To obtain accurate spectra, we recorded absorptions of 3 concentrations of oleylamine in TCE (100% OAm, 10% OAm and 1% OAm), scaled the absorptions of the diluted samples, and stitched the overlapping areas. Spectra of different concentrations were also recorded in the 1.6 μm - 10 μm range using an FTIR spectrometer.

For dodecanethiol and water, we recorded spectra in a 1 cm cuvette and 1 mm cuvette (absorption scaled 10x), and stitched the overlapping areas.

The measured absorbance $A$ for a 1 cm cell was used to calculate the absorption coefficient $\alpha$ by:

$$\alpha = \frac{A \ln(10)}{l} = 2.303\ A\ cm^{-1}$$

The imaginary index of refraction $\epsilon''$ was calculated by:

$$\epsilon'' = \frac{\alpha n \lambda}{2\pi}$$

Where $n$ is the real refractive index and $\lambda$ is the vacuum wavelength. The data are presented in Fig. S3C.

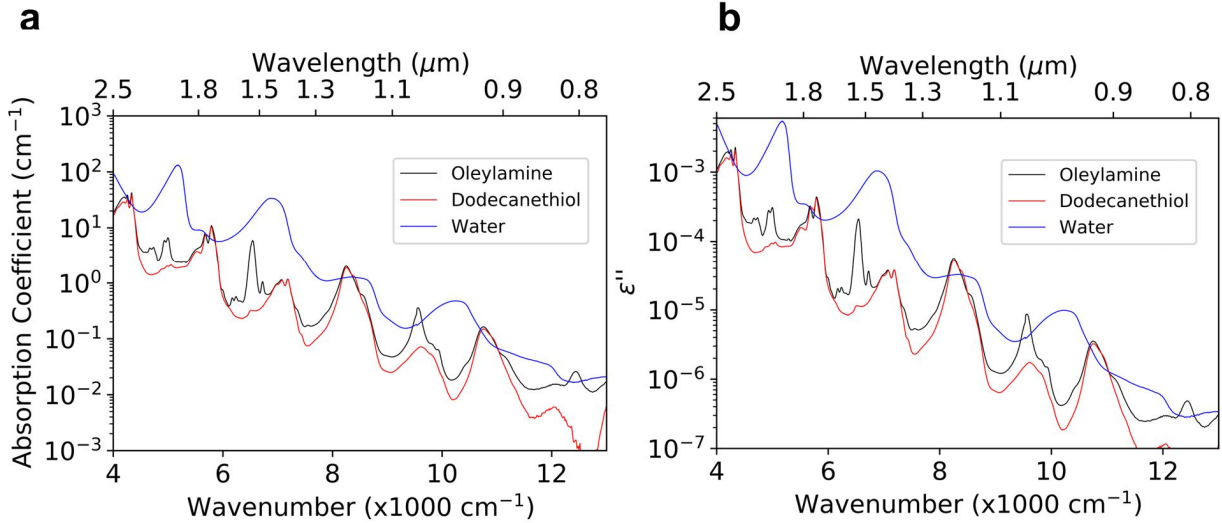

**Fig. S3C**: **a**, Absorption coefficient and **b**, imaginary refractive index of oleylamine, dodecanethiol and water.

Fig. S3C shows that the absorption from oleylamine and dodecanethiol are similar, and mainly determined by the C-H overtones. The absorption from water is nearly 5 times larger than the organic solvents.

### 3D: Calculation of FRET-limited PLQY

The rate of Forster Resonance Energy Transfer (FRET) from the QD to the ligands or solvent is well-described in the supplementary information of Liu. Et. al.[8] and Melnychuk.[9] The nonradiative FRET rate $\gamma_{NR}$ is directly proportional to the radiative rate $\gamma_R$, and the ratio is calculated to be (Eq. 1.3 in Melnychuk)[9]:

$$\frac{\gamma_{NR}}{\gamma_R} = \frac{3\epsilon''}{32\pi^3 n \bar{v}^3} \int_R^{R+\Delta R} \frac{dr}{r^4} \qquad (S3D-1)$$

Where $\epsilon''$ and $n$ are the imaginary and real indices of refraction for the ligand / solvent, $\bar{v}$ is the wavenumber, $R$ is the total nanocrystal radius and $\Delta R$ is the thickness of the ligand shell. Eq. S3D-1 was performed at a single frequency. For a QD with a given absorption spectrum, the ratio of nonradiative to radiative rates can be calculated as:

$$\frac{\gamma_{NR}}{\gamma_R} = \frac{3}{32\pi^3 n} \int_R^{R+\Delta R} \frac{dr}{r^4} \int \frac{\epsilon''(\bar{v}) f_D(\bar{v}) d\bar{v}}{\bar{v}^3} \qquad (S3D-2)$$

Where $f_D(\bar{v})$ is the normalized QD emission spectrum ('D' stands for 'donor', using the terminology of FRET literature).

The photoluminescence quantum yield (PLQY) could be calculated by:

$$\text{PLQY} = \frac{\gamma_R}{\gamma_R + \gamma_{NR}} = \frac{1}{1 + \frac{\gamma_{NR}}{\gamma_R}} \qquad (S3D-3)$$

The magnitude of $\epsilon''$ depends on the surface density of ligands on the QD surface. We used the surface density of oleylamine ligands on a CdSe surface as 1.8 nm$^{-2}$ following the report by Owen and coworkers.[10] To determine the $\epsilon''$ for a surface density of 1.8 nm$^{-2}$, we estimated the surface density if the oleylamine ligand shell had the volume density of bulk oleylamine. Using the density of oleylamine as 0.813 g/cm$^3$, molar mass as 267.5 g/mol, and oleylamine ligand length = 1.8 nm,[8] we estimated the surface coverage to be 4.3 nm$^{-2}$ on a QD of diameter 13 nm. Hence, we scaled the bulk oleylamine $\epsilon''$ by 0.18/4.3 to account for the reported surface coverage.

To calculate the PLQY for QDs in organic solvents capped with oleylamine, $\Delta R$ was set to be 1.8 nm (following Liu et. al.)[8]. For QDs in water, $\Delta R$ was set to infinity. The QD emission spectrum was simulated as a Gaussian with standard deviation as 0.1 times the emission frequency, which was a good approximation for HgSe QDs emitting at 5 μm (intraband) and 2 μm (interband).

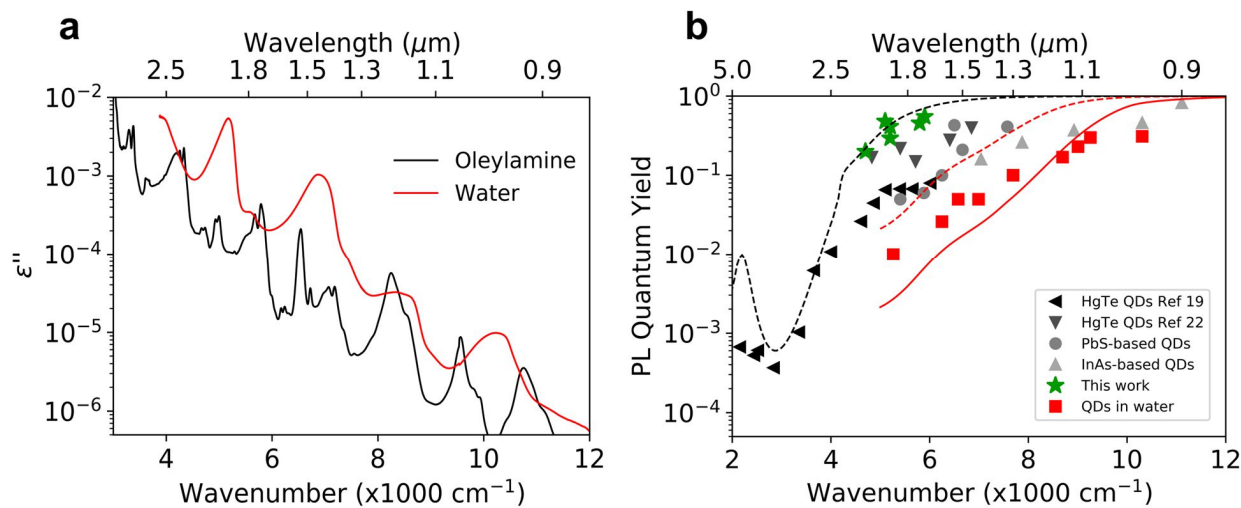

**Fig. S3D**: **a**, Measured imaginary refractive index of oleylamine and water at a function of frequency. **b**, Compilation of PLQYs of the brightest reported QDs at different emission wavelengths (see main text for references and details). The curves are calculated PLQY by FRET to oleylamine ligands (black), and water (red). Solid curve is for 6 nm diameter QDs, and dashed curve is for 13 nm diameter QDs.

# 4. Additional data

## 4A: TEM, SAXS, absorption, PL, and lifetimes of HgSe/CdSe QDs emitting at 2.0 μm

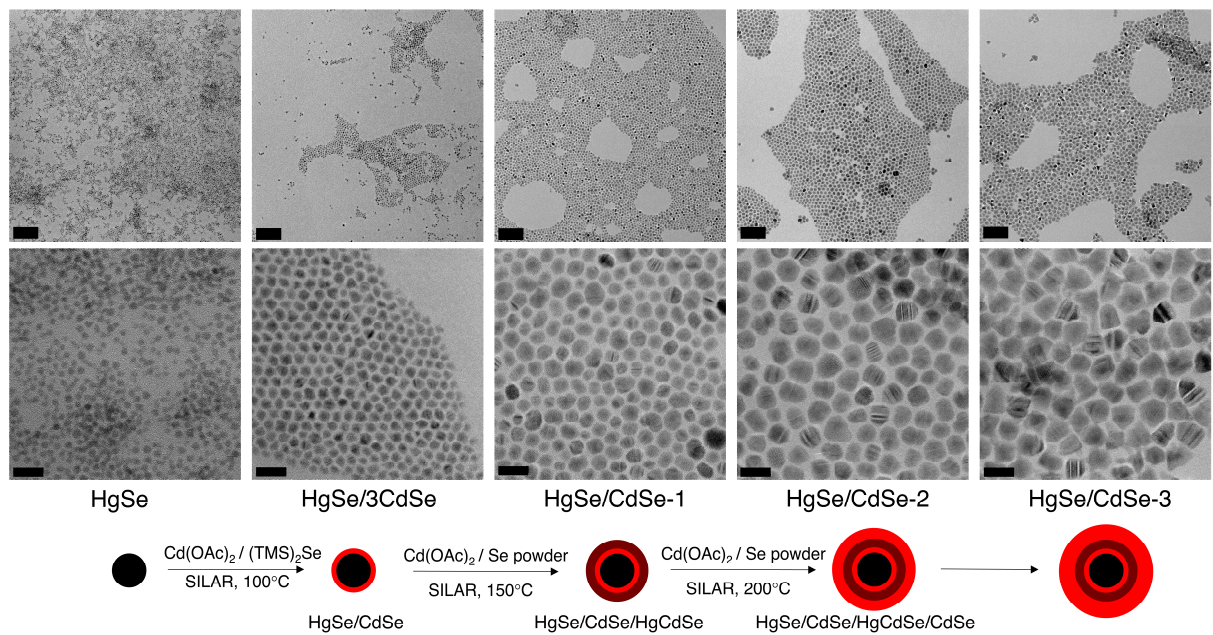

**Fig. S4A-1**: TEM images of HgSe and HgSe/CdSe QDs emitting at 2 μm. The scale bars are 100 nm for the top panels, and 20 nm for the bottom panels.

| Sample | TEM Size (nm) | SAXS Size (nm) |
|---|---|---|
|  |  |  |
| HgSe | 4 ± 0.8 | 5.2 ± 1.2 |
| HgSe/3CdSe | 6.4 ± 0.7 | 7.0 ± 1.1 |
| HgSe/CdSe-1 | 9.6 ± 1.6 | 10.2 ± 1.7 |
| HgSe/CdSe-2 | 13.0 ± 1.8 | 13.0 ± 2.1 |
| HgSe/CdSe-3 | 15.3 ± 1.8 | 15.1 ± 3.4 |

**Table S4A-2**: Sizes of HgSe and HgSe/CdSe QDs measured by SAXS and TEM. The measurement of QD size by TEM were not precise at smaller sizes, and hence the SAXS sizes were used in the main text.

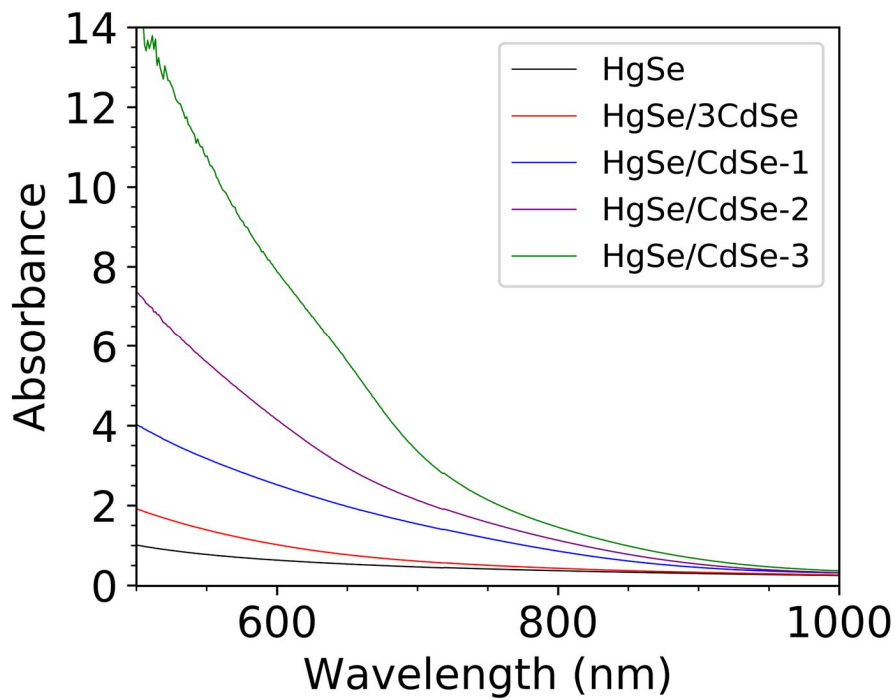

**Fig. S4A-3**: Vis-NIR absorption of HgSe and HgSe/CdSe QDs

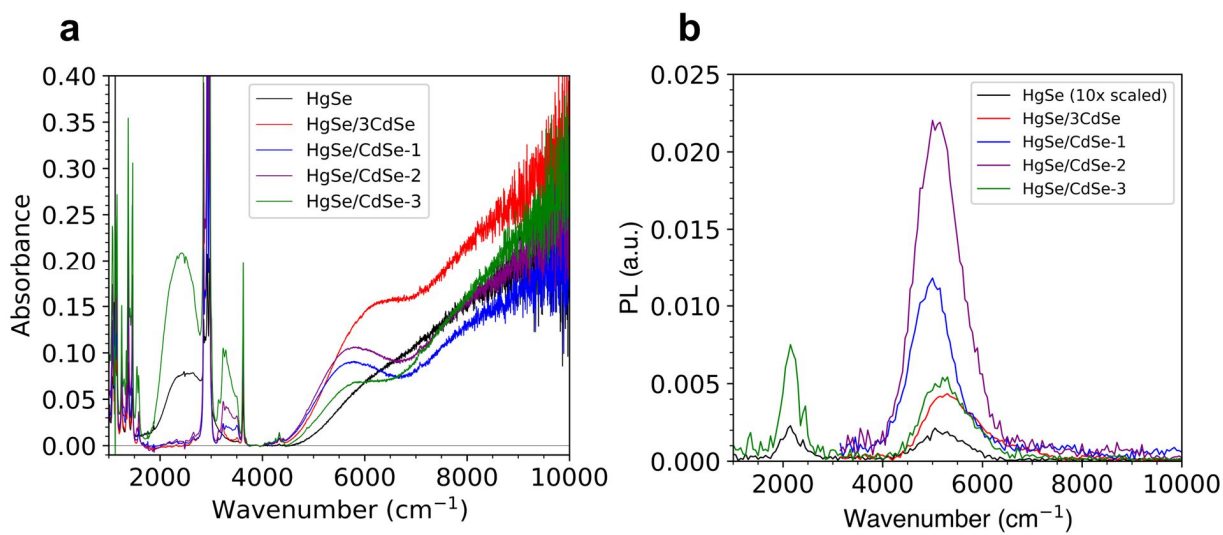

**Fig. S4A-4**: **a**, Absorption spectra of HgSe and HgSe/CdSe QDs. Spectra are quantitative, corresponding to equal masses of cores (as estimated from aliquots). Spectra are vertically shifted to set absorbance = 0 at 3800 cm$^{-1}$. **b**, PL spectra of HgSe and HgSe/CdSe QDs. The HgSe and HgSe/CdSe-3 samples are partially n-doped, and show both intraband and interband PL.

| Sample | PLQE (150mW) | PLQE (15mW) | 1S$_e$ occupancy | PLQY (15mW) |
|---|---|---|---|---|
| HgSe * | 0.4% * | - | 1.04 | 1.8% * |
| HgSe/3CdSe | 14% | 18% | 0 | 18% |
| HgSe/CdSe-1 | 20% | 22% | 0 | 22% |
| HgSe/CdSe-2 | 31% | 36% | 0 | 36% |
| HgSe/CdSe-3 | 13% | 16% | 0.85 | 48% |

**Table S4A-5**: PL quantum efficiency (PLQE) of HgSe and HgSe/CdSe QDs measured using an integrating sphere with and 808 nm excitation, with powers 150 mW and 15 mW. The PLQE was lower at 150 mW excitation, likely due to localized heating of the sample. The PL signal from the HgSe sample was comparable to the PL signal from the blank sample, so the PLQE was estimated using area under the PL spectrum. The doping of HgSe and HgSe/CdSe-3 were determined using the absorption spectrum (see Section 3B), and was used to calculate the PLQY.

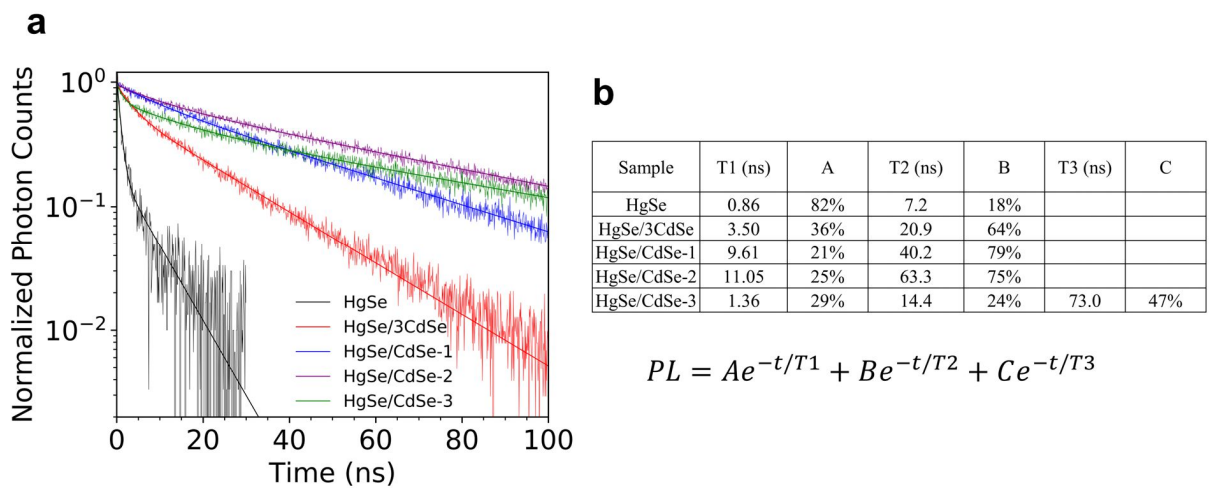

| Sample | T1 (ns) | A | T2 (ns) | B | T3 (ns) | C |
|---|---|---|---|---|---|---|
| HgSe | 0.86 | 82% | 7.2 | 18% | | |
| HgSe/3CdSe | 3.50 | 36% | 20.9 | 64% | | |
| HgSe/CdSe-1 | 9.61 | 21% | 40.2 | 79% | | |
| HgSe/CdSe-2 | 11.05 | 25% | 63.3 | 75% | | |
| HgSe/CdSe-3 | 1.36 | 29% | 14.4 | 24% | 73.0 | 47% |

$$PL = Ae^{-t/T1} + Be^{-t/T2} + Ce^{-t/T3}$$

**Fig. S4A-6**: **a**, PL lifetime data for HgSe and HgSe/CdSe QDs. Data are the same as in Fig. 3b of main text. The HgSe/CdSe-3 sample was fit to a triexponential, while the remaining spectra fit well to biexponential functions. **b**, Fit results.

## 4B: TEM, SAXS, absorption, PL, and lifetimes of HgSe/CdSe QDs emitting at 1.7 μm

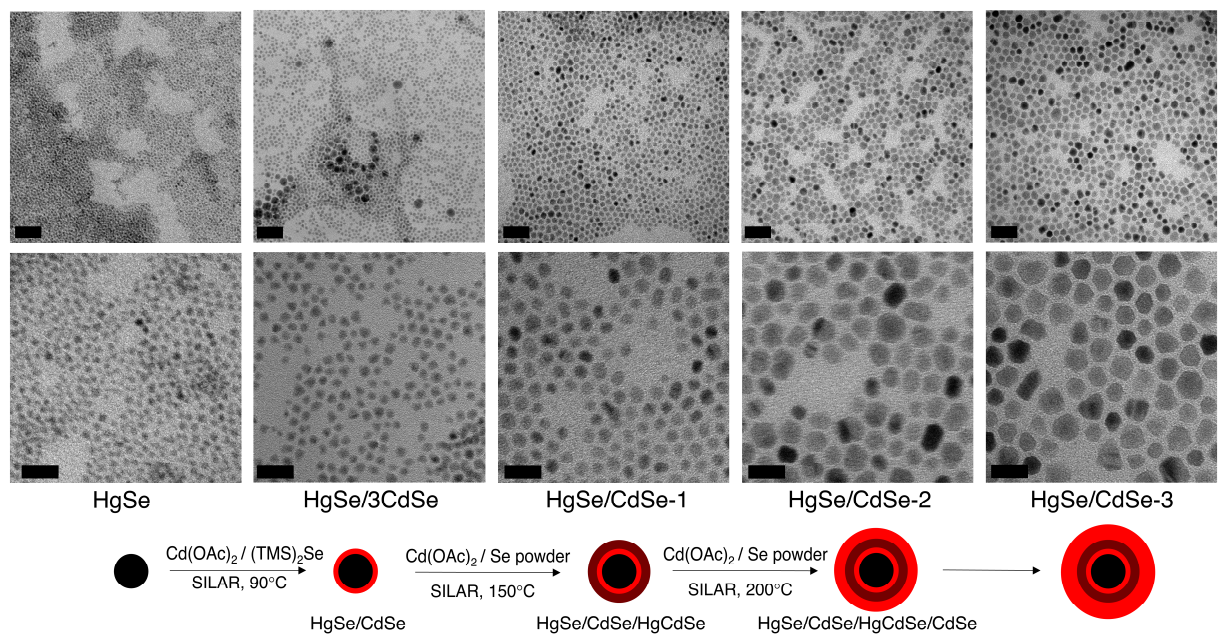

**Fig. S4B-1**: TEM images of HgSe and HgSe/CdSe QDs emitting at 1.7 μm. The scale bars are 50 nm for the top panels, and 20 nm for the bottom panels. There was a broad distribution of QDs seen in the HgSe and HgSe/3CdSe samples, which led to broad PL spectra.

| Sample | TEM Size (nm) | SAXS Size (nm) |
|---|---|---|
|  |  |  |
| HgSe/3CdSe | 5.9 ± 0.8 | 6.3 ± 0.9 |
| HgSe/CdSe-1 | 8.4 ± 1.2 | 9.2 ± 1.8 |
| HgSe/CdSe-2 | 11.7 ± 2.3 | 11.7 ± 2.2 |
| HgSe/CdSe-3 | 12.6 ± 2.4 | 13.3 ± 2.3 |

**Table S4B-2**: Sizes of HgSe and HgSe/CdSe QDs measured by SAXS and TEM.

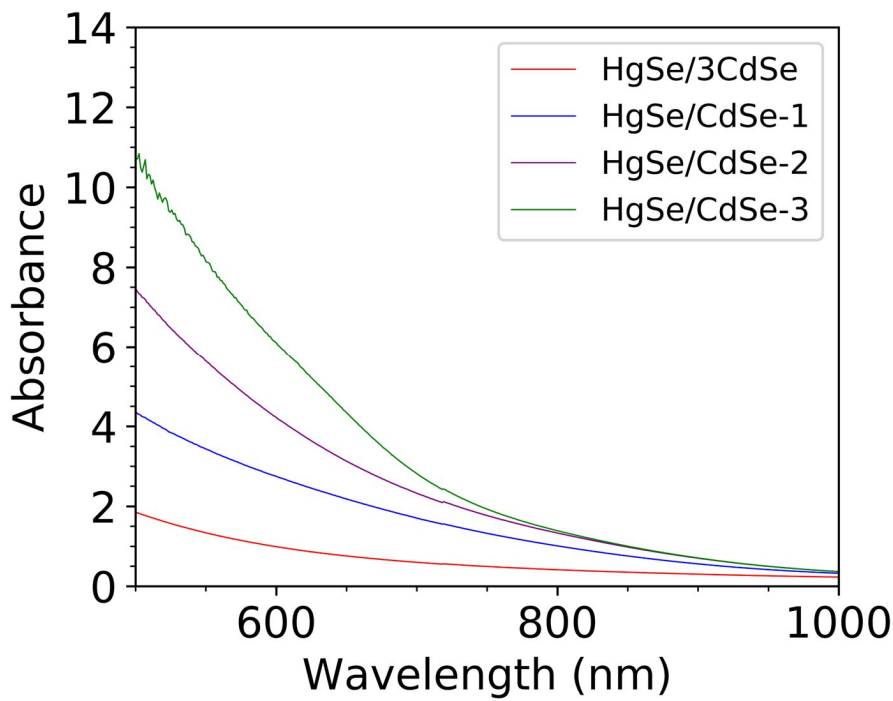

**Fig. S4B-3**: Vis-NIR absorption of HgSe/CdSe QDs

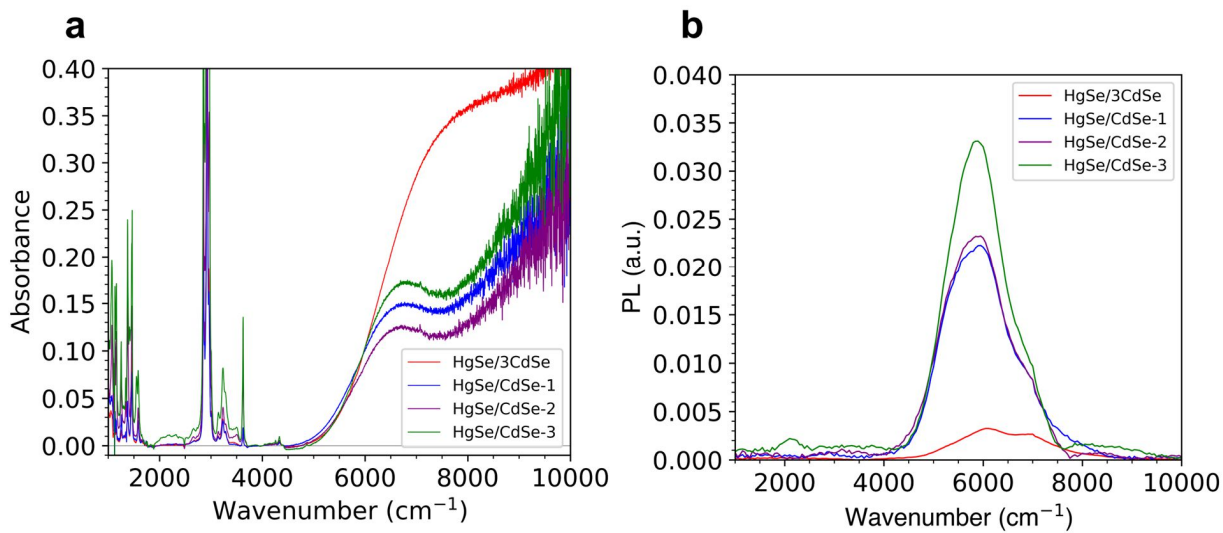

**Fig. S4B-4**: **a**, Absorption spectra of HgSe/CdSe QDs. Spectra are quantitative, corresponding to equal masses of cores (as estimated from aliquots). Spectra are vertically shifted to set absorbance = 0 at 3800 cm$^{-1}$. **b**, PL spectra of HgSe/CdSe QDs. The HgSe/CdSe-3 sample is partially n-doped, and show both intraband and interband PL. The spectra show broad PL due to a broad size distribution of cores.

| Sample | PLQE (150mW) | PLQE (15mW) | 1Se occupancy | PLQY (15mW) |
|---|---|---|---|---|
| HgSe/3CdSe | 14% | 17% | 0 | 17% |
| HgSe/CdSe-1 | 34% | 39% | 0 | 39% |
| HgSe/CdSe-2 | 34% | 43% | 0 | 43% |
| HgSe/CdSe-3 | 44% | 55% | ~0 | 55% |

**Table S4B-5**: PL quantum efficiency (PLQE) of HgSe and HgSe/CdSe QDs measured using an integrating sphere with and 808 nm excitation, with powers 150 mW and 15 mW. The PLQE was lower at 150 mW excitation, likely due to localized heating of the sample.

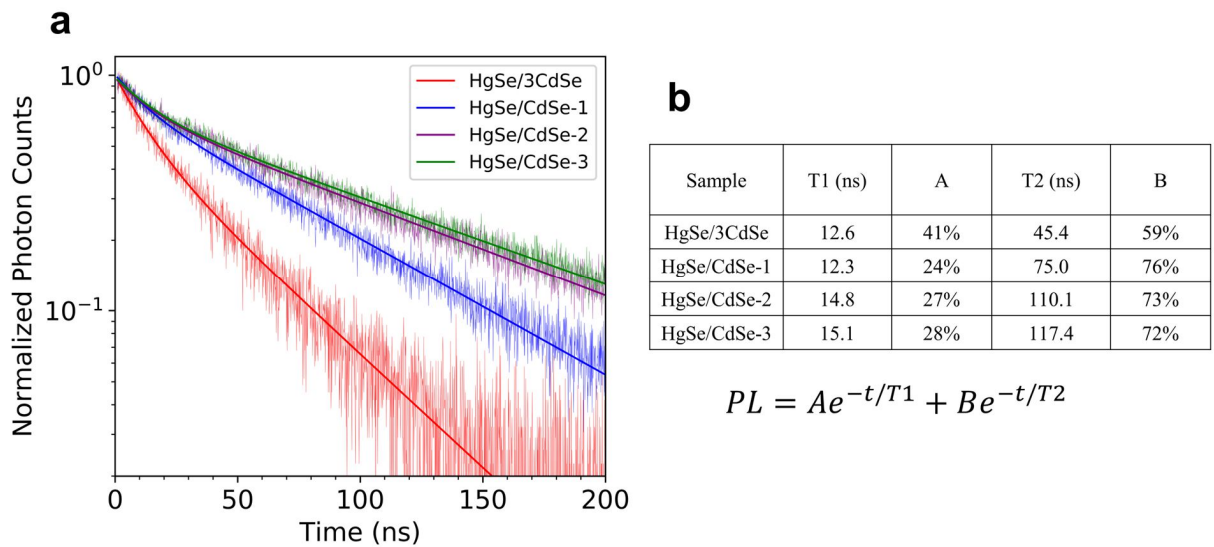

| Sample | T1 (ns) | A | T2 (ns) | B |
|---|---|---|---|---|
| HgSe/3CdSe | 12.6 | 41% | 45.4 | 59% |
| HgSe/CdSe-1 | 12.3 | 24% | 75.0 | 76% |
| HgSe/CdSe-2 | 14.8 | 27% | 110.1 | 73% |
| HgSe/CdSe-3 | 15.1 | 28% | 117.4 | 72% |

$$PL = Ae^{-t/T1} + Be^{-t/T2}$$

**Fig. S4B-6**: **a**, PL lifetime data for HgSe/CdSe QDs. All spectra fit well to biexponential functions. **b**, Fit results.

# 1.7 μm PLQE Measurements

| Sample | $N_{si}$ (mW) | $N_{PbSe}$ (V) | $N_{PbSe-B}$ (V) | Abs (from Si) % | PbSe PL signal (V) | PbSe 808nm signal (V) | Abs (from PbSe) % | Avg abs (%) | PLQE (N) |
|---|---|---|---|---|---|---|---|---|---|
| HgSe/3CdSe-150mW | 0.78 | 6.2 | 0.61 | 59% | 1.08 | 5.12 | 61% | 60% | 13.7% |
| HgSe/3CdSe-15mW | 0.078 | 0.65 | 0.076 | 59% | 0.13 | 0.52 | 61% | 60% | 17.1% |
| HgSe/CdSe-1-150mW | 0.814 | 7.44 | 1.48 | 57% | 2.62 | 4.82 | 63% | 60% | 33.0% |
| HgSe/CdSe-1-15mW | 0.082 | 0.82 | 0.168 | 56% | 0.30 | 0.52 | 60% | 58% | 38.6% |
| HgSe/CdSe-2-150mW | 0.972 | 8.08 | 1.28 | 48% | 2.27 | 5.81 | 56% | 52% | 32.9% |
| HgSe/CdSe-2-15mW | 0.098 | 0.88 | 0.162 | 48% | 0.29 | 0.59 | 55% | 51% | 42.2% |
| HgSe/CdSe-3-150mW | 0.915 | 8.24 | 1.8 | 51% | 3.19 | 5.05 | 62% | 57% | 42.7% |
| HgSe/CdSe-3-15mW | 0.091 | 0.94 | 0.224 | 52% | 0.40 | 0.54 | 59% | 55% | 54.4% |
| TCE 150mW | 1.88 | 13.2 | 0.0195 | | | | | | |

| Sample | $O_{si}$ (mW) | $O_{PbSe}$ (V) | $O_{PbSe-B}$ (V) | Abs (from Si) % | PbSe PL signal (V) | PbSe 808nm signal | Abs (from PbSe) % | Avg abs (%) | PLQE (O) |
|---|---|---|---|---|---|---|---|---|---|
| HgSe/3CdSe-150mW | 0.833 | 6.56 | 0.656 | 56% | 1.16 | 5.40 | 61% | 58% | 14.4% |
| HgSe/3CdSe-15mW | 0.083 | 0.688 | 0.08 | 56% | 0.14 | 0.55 | 60% | 58% | 17.7% |
| HgSe/CdSe-1-150mW | 0.878 | 7.76 | 1.58 | 53% | 2.80 | 4.96 | 64% | 59% | 34.5% |
| HgSe/CdSe-1-15mW | 0.088 | 0.848 | 0.178 | 53% | 0.32 | 0.53 | 61% | 57% | 39.9% |
| HgSe/CdSe-2-150mW | 1.05 | 8.64 | 1.39 | 44% | 2.46 | 6.18 | 55% | 50% | 35.9% |
| HgSe/CdSe-2-15mW | 0.106 | 0.944 | 0.168 | 44% | 0.30 | 0.65 | 53% | 48% | 44.5% |
| HgSe/CdSe-3-150mW | 0.976 | 8.8 | 1.91 | 48% | 3.38 | 5.42 | 61% | 54% | 45.0% |
| HgSe/CdSe-3-15mW | 0.098 | 0.96 | 0.23 | 48% | 0.41 | 0.55 | 60% | 54% | 54.7% |
| TCE 150mW | 2.02 | 13.8 | 0.021 | | | | | | |

| Sample | PLQE (150mW) | Error | PLQE (15mW) | Error | 15/150 Ratio |
|---|---|---|---|---|---|
| HgSe/3CdSe | 14% | 0.039 | 17% | 0.024 | 1.24 |
| HgSe/CdSe-1 | 34% | 0.032 | 39% | 0.023 | 1.16 |
| HgSe/CdSe-2 | 34% | 0.061 | 43% | 0.037981 | 1.26 |
| HgSe/CdSe-3 | 44% | 0.037 | 55% | 0.004 | 1.24 |

PLQY error avg  0.032

## 2.0 μm PLQE Measurements

| Sample | $N_{si}$ (mW) | $N_{PbSe}$ (V) | $N_{PbSe-B}$ (V) | Abs (from Si) % | PbSe PL signal (V) | PbSe 808nm signal (V) | Abs (from PbSe) % | Avg abs (%) | PLQE (N) |
|---|---|---|---|---|---|---|---|---|---|
| HgSe-150mW | 0.743 | 4.96 | 0.032 | 60% | 0.06 | 4.90 | 57% | 59% | 0.8% |
| HgSe/3CdSe-150mW | 0.764 | 5.68 | 0.528 | 59% | 0.93 | 4.75 | 59% | 59% | 13.8% |
| HgSe/3CdSe-15mW | 0.075 | 0.58 | 0.068 | 59% | 0.12 | 0.46 | 60% | 60% | 17.5% |
| HgSe/CdSe-1-150mW | 0.602 | 5.2 | 0.904 | 67% | 1.60 | 3.60 | 69% | 68% | 20.4% |
| HgSe/CdSe-1-15mW | 0.06 | 0.54 | 0.099 | 68% | 0.18 | 0.36 | 68% | 68% | 22.4% |
| HgSe/CdSe-2-150mW | 0.307 | 4.4 | 1.72 | 83% | 3.04 | 1.36 | 88% | 86% | 30.8% |
| HgSe/CdSe-2-15mW | 0.031 | 0.484 | 0.2 | 83% | 0.35 | 0.13 | 89% | 86% | 35.8% |
| HgSe/CdSe-3-150mW | 0.602 | 4.72 | 0.572 | 67% | 1.01 | 3.71 | 68% | 68% | 13.0% |
| HgSe/CdSe-3-15mW | 0.06 | 0.488 | 0.0688 | 68% | 0.12 | 0.37 | 68% | 68% | 15.6% |
| TCE 150mW | 1.85 | 11.5 | 0.025 | | | | | | |

| Sample | $O_{si}$ (mW) | $O_{PbSe}$ (V) | $O_{PbSe-B}$ (V) | Abs (from Si) % | PbSe PL signal (V) | PbSe 808nm signal | Abs (from PbSe) % | Avg abs (%) | PLQE (O) |
|---|---|---|---|---|---|---|---|---|---|
| HgSe-150mW | 0.796 | 5.16 | 0.032 | 57% | 0.06 | 5.10 | 58% | 58% | 0.8% |
| HgSe/3CdSe-150mW | 0.827 | 6 | 0.56 | 55% | 0.99 | 5.01 | 59% | 57% | 14.2% |
| HgSe/3CdSe-15mW | 0.08 | 0.612 | 0.072 | 57% | 0.13 | 0.48 | 60% | 59% | 17.8% |
| HgSe/CdSe-1-150mW | 0.647 | 5.4 | 0.936 | 65% | 1.66 | 3.74 | 69% | 67% | 20.2% |
| HgSe/CdSe-1-15mW | 0.065 | 0.56 | 0.104 | 65% | 0.18 | 0.38 | 69% | 67% | 22.5% |
| HgSe/CdSe-2-150mW | 0.366 | 4.72 | 1.8 | 80% | 3.19 | 1.53 | 87% | 84% | 31.2% |
| HgSe/CdSe-2-15mW | 0.037 | 0.524 | 0.206 | 80% | 0.36 | 0.16 | 87% | 83% | 35.8% |
| HgSe/CdSe-3-150mW | 0.66 | 4.96 | 0.584 | 64% | 1.03 | 3.93 | 68% | 66% | 12.8% |
| HgSe/CdSe-3-15mW | 0.066 | 0.532 | 0.0724 | 64% | 0.13 | 0.40 | 67% | 66% | 16.0% |
| TCE 150mW | 1.98 | 12.2 | 0.026 | | | | | | |

| Sample | PLQE (150mW) | Error | PLQE (15mW) | Error | 15/150 Ratio |
|---|---|---|---|---|---|
| HgSe | 0.8% | 0.029 | | | |
| HgSe/3CdSe | 14% | 0.019 | 18% | 0.013 | 1.26 |
| HgSe/CdSe-1 | 20% | 0.008 | 22% | 0.002 | 1.11 |
| HgSe/CdSe-2 | 31% | 0.007 | 36% | 0.000001 | 1.15 |
| HgSe/CdSe-3 | 13% | 0.011 | 16% | 0.018 | 1.22 |

PLQY error avg     0.012

# Reported PLQY Values

## Data in green rows are included in Fig. 4b in the main text

| Link | Reference | Material | Wavelength (nm) | Wavenumber (cm$^{-1}$) | Solvent | PLQY |
|---|---|---|---|---|---|---|
| https://onlinelibrary.wiley.com/doi/full/10.1002/advs.202200637 | Konstantatos Adv Mater 2022 | PbS film structure, MPA coated | 1538 | 6500 | Film | 65% |
| https://www.nature.com/articles/s41565-018-0312-y | Konstantatos Nat Nanotech 2019 | PbS film structure, MPA coated | 1408 | 7100 | Film | 80% |
| https://pubs.acs.org/doi/10.1021/jz100830r | Beard JPCL 2010 | PbS QDs, oleate capped | 1320 | 7576 | Nonpolar | 41% |
| | Beard JPCL 2010 | PbS QDs, oleate capped | 1500 | 6667 | Nonpolar | 21% |
| | Beard JPCL 2010 | PbS QDs, oleate capped | 1600 | 6250 | Nonpolar | 10% |
| | Beard JPCL 2010 | PbS QDs, oleate capped | 1700 | 5882 | Nonpolar | 6% |
| | Beard JPCL 2010 | PbS QDs, oleate capped | 1850 | 5405 | Nonpolar | 5% |
| https://www.nature.com/articles/s41566-019-0526-z | Mohd Yusoff Nat Photon 2020 | Ag2S-SiO2 QDs in perovskite matrix | 1400 | 7143 | Film | 84% |
| https://pubs.rsc.org/en/content/articlelanding/2014/ra/c4ra06098a | Takashi Jin, RSC Adv 2014 | PbS/CdS, capped with MUA | 1150 | 8696 | Water | 17% |
| | Takashi Jin, RSC Adv 2014 | PbS, capped with MUA | 1100 | 9091 | Water | 8% |
| | Takashi Jin, RSC Adv 2014 | PbS/CdS, OAm/oleate | 1150 | 8696 | CHCl3 | 44% |
| | Takashi Jin, RSC Adv 2014 | PbS, OAm/oleate | 1100 | 9091 | CHCl3 | 37% |
| https://pubs.acs.org/doi/full/10.1021/acsnano.0c05907 | Hollingsworth ACS Nano 2021 | PbS/CdS thin shell | 1550 | 6500 | Nonpolar | 43% |
| https://onlinelibrary.wiley.com/doi/full/10.1002/smll.202001003 | Qiangbin Wang, Small 2020 | Ag2Te/Ag2S QDs, DDT-acetate capped | 1300 | 7692 | Nonpolar | 0.43% |
| https://www.nature.com/articles/ncomms12749 | Bawendi, Nature Comm 2016 | InAs/CdSe/CdS, OAm-oleate capped | 1420 | 7042 | Nonpolar | 16% |
| | Bawendi, Nature Comm 2016 | InAs/CdSe/CdS, OAm-oleate capped | 1270 | 7874 | Nonpolar | 26% |
| | Bawendi, Nature Comm 2016 | InAs/CdSe/CdS, OAm-oleate capped | 1120 | 8929 | Nonpolar | 37% |
| | Bawendi, Nature Comm 2016 | InAs/CdSe/CdS, OAm-oleate capped | 900 | 11111 | Nonpolar | 82% |
| | | | 970 | 10309 | Nonpolar | 46% |
| | | | 970 | 10309 | H2O | 31% |
| | | | 1110 | 9009 | H2O | 23% |
| | | | 1300 | 7692 | H2O | 10% |
| | | | 1430 | 6993 | H2O | 5% |
| https://www.nature.com/articles/s41551-017-0056#Sec35 | Bawendi Nature Biomed Engg 2017 | InAs/CdSe/ZnSe | 1080 | 9259 | H2O | 30% |
| | | | 1280 | 7813 | H2O | 4% |
| https://www.nature.com/articles/s41587-019-0262-4 | Hongjie Dai Nature Biotech 2019 | NaYbF4:Ce,Er,Zn/NaYF4, PEG-coated | 1520 | 6579 | H2O | 5% |
| https://www.pnas.org/doi/10.1073/pnas.1806153115 | Hongjie Dai PNAS 2018 | PbS/CdS, OAm-polyacrylicacid - PEG | 1650 | 6061 | H2O | 2.20% |
| | Ben Zhong Tang Nat Comm 2020 | AIEgens | 1100 | 9091 | H2O | 1.15% |
| https://onlinelibrary.wiley.com/doi/full/10.1002/adma.202103953 | Qiangbin Wang, Adv Mater 2021 | Au-Ag2Te | 1490 | 6711 | Nonpolar | 6.20% |
| https://www.nature.com/articles/s41565-022-01130-3 | Hongjie Dai Nature Nanotech 2022 | PbS/CdS PEGylated | 1900 | 5263 | H2O | 1% |
| | | | 1600 | 6250 | H2O | 2.60% |
| https://pubs.acs.org/doi/full/10.1021/acs.jpclett.0c00958 | Justin Caram JPCL 2020 | HgTe NPs | 1500 | 6667 | Nonpolar | 15% |

| Link | Reference | Material | Wavelength (nm) | Wavenumber (cm$^{-1}$) | Solvent | PLQY |
|---|---|---|---|---|---|---|
| https://pubs.acs.org/doi/full/10.1021/jp409061g | Keuleyan JPCC 2014 | HgTe QDs | 4777 | 2093 | Nonpolar | 1.96E-04 |
| | | | 4642 | 2154 | Nonpolar | 6.70E-04 |
| | | | 4515 | 2215 | Nonpolar | 3.25E-04 |
| | | | 4254 | 2351 | Nonpolar | 1.78E-04 |
| | | | 4097 | 2441 | Nonpolar | 5.25E-04 |
| | | | 3971 | 2518 | Nonpolar | 6.08E-04 |
| | | | 3747 | 2669 | Nonpolar | 1.61E-04 |
| | | | 3603 | 2776 | Nonpolar | 1.73E-04 |
| | | | 3603 | 2776 | Nonpolar | 2.43E-04 |
| | | | 3526 | 2836 | Nonpolar | 3.66E-04 |
| | | | 2986 | 3349 | Nonpolar | 5.39E-04 |
| | | | 2986 | 3349 | Nonpolar | 0.00103 |
| | | | 2738 | 3653 | Nonpolar | 0.00616 |
| | | | 2693 | 3713 | Nonpolar | 7.55E-04 |
| | | | 2500 | 4000 | Nonpolar | 0.01075 |
| | | | 2435 | 4107 | Nonpolar | 0.00484 |
| | | | 2171 | 4607 | Nonpolar | 0.0262 |
| | | | 2136 | 4682 | Nonpolar | 0.00804 |
| | | | 2056 | 4864 | Nonpolar | 0.04468 |
| | | | 2024 | 4940 | Nonpolar | 0.02443 |
| | | | 1959 | 5105 | Nonpolar | 0.06557 |
| | | | 1854 | 5394 | Nonpolar | 0.06734 |
| | | | 1769 | 5651 | Nonpolar | 0.06734 |
| | | | 1663 | 6013 | Nonpolar | 0.07956 |
| | | | 1576 | 6346 | Nonpolar | 0.06893 |
| https://pubs.acs.org/doi/full/10.1021/acs.chemmater.7b02637 | Rogach Chem Mater 2017 | HgTe | 2070 | 4831 | Nonpolar | 0.17 |
| | | | 1850 | 5405 | Nonpolar | 0.22 |
| | | | 1460 | 6849 | Nonpolar | 0.4 |
| | | | 1560 | 6410 | Nonpolar | 0.28 |
| | | | 1750 | 5714 | Nonpolar | 0.15 |
| | This work | HgSe/CdSe, OAm, acetate capped | 1961 | 5100 | Nonpolar | 0.48 |
| | | | 1695 | 5900 | Nonpolar | 0.55 |
| | | | 1724 | 5800 | Nonpolar | 0.45664 |
| | | | 1923 | 5200 | Nonpolar | 0.41 |
| | | | 1923 | 5200 | Nonpolar | 0.29 |
| | | | 2128 | 4700 | Nonpolar | 0.20 |